\documentclass[apj,numberedappendix]{emulateapj}
\usepackage{graphicx}
\usepackage{grffile}
\usepackage{subfigure}
\usepackage{mathrsfs,amssymb}
\usepackage{amsmath}
\usepackage{natbib}
\usepackage{color}
\usepackage{ulem}
\usepackage{url}
\usepackage{bm}
\usepackage{multirow}
\usepackage{gensymb}
\usepackage{accents}

\newcommand{\varA}[1]{{\operatorname{#1}}}

\newcommand*{\dt}[1]{\accentset{\mbox{\large\bfseries .}}{#1}}

\def\gtsima{$\; \buildrel > \over \sim \;$}
\def\ltsima{$\; \buildrel < \over \sim \;$}
\def\gsim{\lower.5ex\hbox{\gtsima}}
\def\lsim{\lower.5ex\hbox{\ltsima}}
\def\simgt{\lower.5ex\hbox{\gtsima}}
\def\simlt{\lower.5ex\hbox{\ltsima}}
\def\simpr{\lower.5ex\hbox{\prosima}}
\def\mean#1{\left< #1 \right>}
 
 \newcommand*\oline[1]{%
  \vbox{%
    \hrule height 0.5pt%                  % Line above with certain width
    \kern0.25ex%                          % Distance between line and content
    \hbox{%
      \kern-0.1em%                        % Distance between content and left side of box, negative values for lines shorter than content
      \ifmmode#1\else\ensuremath{#1}\fi%  % The content, typeset in dependence of mode
      \kern-0.1em%                        % Distance between content and left side of box, negative values for lines shorter than content
    }% end of hbox
  }% end of vbox
}

\begin{document}
 
\title{Triggering the formation of direct collapse black holes by their congeners}
\author{Bin Yue$^{1, 2}$, Andrea Ferrara$^{2, 3}$,   Fabio Pacucci$^{2,4}$, Kazuyuki Omukai$^{5}$} 
\affil{$^1$National Astronomical Observatories, Chinese Academy of Sciences, Beijing, 100012, China} 
\affil{$^2$Scuola Normale Superiore, Piazza dei Cavalieri 7, I-56126 Pisa, Italy}
\affil{$^3$Kavli IPMU (WPI), Todai Institutes for Advanced Study, the University of Tokyo, Japan}
\affil{$^4$Department of Physics, Yale University, P.O. Box 208101, New Haven, CT 06520, USA}
\affil{$^5$Astronomical Institute, Tohoku University, Sendai, Miyagi 980-8578, Japan}
\email{yuebin@bao.ac.cn}
 
\begin{abstract}
Direct collapse black holes (DCBHs) are excellent candidates as seeds of supermassive black holes (SMBHs) observed at $z \gsim 6$. The formation of a DCBH requires a strong external radiation field to suppress $\rm H_2$ formation and cooling in a collapsing gas cloud. Such strong field is not easily achieved by first stars or normal star-forming galaxies. Here we investigate a scenario in which the previously-formed DCBH can provide the necessary radiation field for the formation of additional ones. Using one-zone model and the simulated DCBH Spectral Energy Distributions (SEDs) filtered through absorbing gas initially having column density $N_{\rm H}$, we derive the critical field intensity, $J_{\rm LW}^{\rm crit}$, to suppress $\rm H_2$ formation and cooling. For the SED model with $N_{\rm H}=1.3\times10^{25}$ cm$^{-2}$, $8.0\times10^{24}$ cm$^{-2}$  and  $5.0\times10^{24}$ cm$^{-2}$, we obtain $J_{\rm LW}^{\rm crit}\approx22$, 35 and 54, all much smaller than the critical field intensity for normal star-forming galaxies ($J_{\rm LW}^{\rm crit}\simgt 1000$). X-ray photons from previously-formed DCBHs build up a high-$z$ X-ray background (XRB) that may boost the $J_{\rm LW}^{\rm crit}$. However, we find that in the three SED models $J_{\rm LW}^{\rm crit}$ only increases to $\approx80$, 170 and 390 respectively even when $\dt{\rho}_\bullet$ reaches the maximum value allowed by the present-day XRB level ($0.22, 0.034, 0.006~M_\odot$yr$^{-1}$Mpc$^{-3}$), still much smaller than the galactic value. Although considering the XRB from first galaxies may further increase $J_{\rm LW}^{\rm crit}$, we conclude that our investigation supports a scenario in which DCBH may be more abundant than predicted by models only including galaxies as external radiation sources.
\end{abstract}


\section{Introduction}

The quest for supermassive black holes (SMBHs) at $z\gsim 6$ has been a remarkable success 
\citep{ 2003AJ....125.1649F,  2007AJ....134.1150J,  2007ApJ...669...32K,   2011Natur.474..616M,  2012AJ....143..142M,  2013ApJ...779...24V,  2015Natur.518..512W}. 
However,  the detection of such SMBHs as massive as $\sim10^9 -10^{10}~M_\odot$ at such early epochs with cosmic age $\lsim 1$ Gyr has opened 
a number of questions that remain essentially unanswered. 

Black holes (BHs) grow from an initial seed by accreting the surrounding gas. 
Naive SMBH seed candidates are stellar-mass BHs formed 
after the death of massive (possibly Pop III) stars, with a typical mass $\sim 100~M_\odot$ or even smaller \citep{2003ApJ...591..288H, 2011Sci...334.1250H}. However, even assuming a continuous accretion at the Eddington limit, the time required to grow a $\sim10^9~M_\odot$ SMBH by $z\sim6$ is comparable or even longer than the Hubble time at that redshift. Even worse, in the early evolutionary stages, radiative feedback reduces the accretion rate significantly. As a result, for most of the time the BH grows slowly \citep{2009ApJ...701L.133A, 2012ApJ...754...34J,2016MNRAS.457.3356V}. Furthermore, reionization of the intergalactic medium (IGM) may also significantly hamper the growth process \citep{2012MNRAS.425.2974T}. 

To overcome these problems, an alternative scenario in which a SMBH grows from a seed more massive than $10^4~M_\odot$ has become popular (see e.g. review \citealt{2010A&ARv..18..279V}). Such massive seeds are nicely produced by the so-called ``direct collapse" process occurring  in metal-free atomic-cooling halos with virial temperature $T_{\rm vir} \gsim 10^4$~K (\citealt{2003ApJ...596...34B,2004MNRAS.354..292K,2006MNRAS.371.1813L,2006MNRAS.370..289B,2008MNRAS.387.1649B,2010MNRAS.402..673B,2009MNRAS.396..343R,2013ApJ...771..116J,2013MNRAS.433.1607L,2013MNRAS.436.2989L,2016MNRAS.458..233L,2016PASA...33...51L,2015MNRAS.446.3163L,2014MNRAS.443.1979L,2016MNRAS.463..529H,2016MNRAS.459.4209A, 2012MNRAS.425.2854A,2013MNRAS.432.3438A,2014MNRAS.443..648A,2015MNRAS.450.4411C, 2014ApJ...795..137R,2015MNRAS.454.3771P,2015MNRAS.452.1922P,2015MNRAS.448..104P}). If in these halos $\rm H_2$ formation and cooling is suppressed, metal-free gas can only cool via Ly$\alpha$ radiation that becomes inefficient when the gas cools below $\sim 8000$ K. Then, the gas contracts almost isothermally with $T\sim 8000$ K and avoids fragmentation during the cloud collapse \citep{2014MNRAS.445L.109I}. Eventually at the center a BH with mass $\sim10^4 - 10^6~M_\odot$ forms \citep{2014MNRAS.443.2410F}. Thereafter such a direct collapse black hole (DCBH) continues to accrete the gas and may merge with other BHs. A fully-fledged SMBH is then produced in time scale much shorter than the Hubble time. 

The suppression of H$_2$ formation requires either a strong Lyman-Werner (LW, $11.2  < h\nu < 13.6$ eV) UV radiation field that directly dissociates H$_2$, or  a strong continuum radiation field from near-infrared (NIR) to UV band ($0.755 < h\nu < 13.6$ eV) that detaches the most important $\rm H_2$ formation catalyst, H$^-$. The required critical field intensity and the clustering of the dark matter halos determine the actual abundance  of DCBHs in high-$z$ Universe \citep{2008MNRAS.391.1961D,2014MNRAS.440.1263Y}.  The DCBH abundance has been studied by many authors and the predicted number density at $z\sim10$ ranges from $\sim 10^{-10}$ Mpc$^{-3}$ to $\sim10^{-3}$ Mpc$^{-3}$ (e.g. \citealt{2012MNRAS.425.2854A,2013MNRAS.432.3438A, 2014MNRAS.442.2036D, 2014MNRAS.445.1056V,  2015MNRAS.454.3771P, 2016MNRAS.463..529H}). Nevertheless such number density is much smaller than that of galaxies in the high-$z$ Universe.

Generally, in this scenario the required critical field intensity is very high.  For an ideal blackbody spectrum with effective temperature $10^4$ K, $J_{\nu_{13.6}}^{\rm crit}\sim 30 - 300$ \citep{2001ApJ...546..635O, 2010MNRAS.402.1249S}; for a realistic galaxy spectrum $J_{\nu_{13.6}}^{\rm crit}\gsim 1000 - 10000$ \citep{2014MNRAS.445..544S,2015MNRAS.446.3163L},
where $J_{\nu_{13.6}}$ is the specific intensity of the radiation field at 13.6 eV and in units $10^{-21}$ erg s$^{-1}$cm$^{-2}$Hz$^{-1}$sr$^{-1}$.
Such a strong radiation field can be attained only when the radiation source is very close to the DCBH-forming gas cloud. 
In spite of this stringent requirement, the discovery of two high-$z$ DCBH candidates in the CANDELS/GOODS-S survey has been recently reported \citep{2016MNRAS.459.1432P}. 
Moreover, the Lyman alpha emitter (LAE) ``CR7" has been observed and reported as a DCBH candidate by some authors,
because its strong Ly$\alpha$ and HeII 1640~\AA~line luminosities are hard to be provided by Pop III stars; 
and the two nearby Pop II galaxies are ideal sources for providing strong external radiation, 
although this detection is still a matter of debate 
(\citealt{2015ApJ...808..139S, 2015MNRAS.453.2465P, 2016MNRAS.460.3143S, 2016MNRAS.460.4003A, 2016ApJ...829L...6S, 2016arXiv160900727B,2017MNRAS.468L..77P}).

In \citet{2014MNRAS.440.1263Y} we proposed that an accreting and Compton-thick DCBH can provide a radiation field that suppresses H$_2$ formation 
in a more efficient way compared with a galaxy. The advantages here are, for the Spectral Energy Distribution (SED) of a Compton-thick DCBH, the ratio of the H$^-$ photo-detachment rate to the H$_2$ photo-dissociation rate (this ratio is  a most straightforward indicator of the H$_2$ suppression efficiency, see, \citealt{2014MNRAS.445..544S,2017MNRAS.tmp..173W}) is higher than for normal star-forming galaxies; and a BH is usually brighter than a galaxy. Moreover, there is an additional advantage: if the source also produces X-ray and ionizing photons, in general it may hamper DCBH formation \citep{2016arXiv161005679G, 2016MNRAS.461..111R, 2014ApJ...797..139A, 2015MNRAS.450.4350I,2014MNRAS.445..686J, 2011MNRAS.416.2748I}; however if the source is a Compton-thick DCBH, except a few very hard X-ray photons (i.e. $\gsim$ 10 keV) during most of the accretion stage it does not emit X-rays and ionizing photons. In this case, the abundance of high-$z$ DCBHs can be higher than currently expected.

In this work, we derive the critical field intensity for the formation of a new DCBH inside a collapsing gas cloud irradiated by a nearby previously-formed DCBH by means of  
the one-zone model in which the whole cloud is assumed to have uniform properties. 
The SEDs of the DCBH are taken from numerical simulations \citep{2015MNRAS.454.3771P}. We consider three SED models with {\it initial} column number densities $N_{\rm H} = 1.3\times10^{25}$ cm$^{-2}$, $N_{\rm H} = 8.0\times10^{24}$ cm$^{-2}$ and $N_{\rm H} = 5.0\times10^{24}$ cm$^{-2}$ respectively. Throughout this paper $N_{\rm H}$ refers to the column number density of the gas envelope that encloses the accretion disc, at the time when accretion just starts. The real column density filtering the BH radiation is time-dependent and decreases as accretion processes.

Regarding the X-ray radiation, although during most of its accretion process the nearby DCBH only emits hard X-ray photons that negligibly influence the H$_2$ formation, the collective X-ray emission\footnote{Generally soft X-ray photons are trapped in the high column density matter and only hard X-ray photons contribute to the XRB. However, as the matter is swallowed by the DCBH the column density decreases, eventually soft X-ray photons can also escape from the BH \citep{2015MNRAS.454.3771P} and contribute to the XRB as well.} from distant DCBHs and galaxies builds up a high-$z$ X-ray background (XRB) that ionizes the gas and may promote the $\rm H_2$ formation.  We also investigate the impact of such high-$z$ XRB on the DCBH formation. 

An external radiation field is not the only factor that can suppress the H$_2$ formation and lead to the formation of a DCBH. \citet{2012MNRAS.422.2539I} pointed out that in a protogalaxy the cold and dense accretion inflows collide with each other near the center, the shocked gas then forms a hot and dense core, where H$_2$ formation would be suppressed due to collisional dissociation. However, \citet{2014MNRAS.442L.100V} argued that the density required for this scenario to occur is unable to be achieved in a realistic halo. \citet{2015MNRAS.453.1692I} however proposed that the high density could be achieved by collision between two protogalaxies with a high relative velocity. On the other hand, \citet{2010Natur.466.1082M} proposed that when two halos merge, a massive and unstable disc could be produced by merger-driven inflows. A central core forms by accretion from this disc with a high rate ($\gsim10^4~\dot{M}_\odot$yr$^{-1}$ ), no matter whether the gas is pristine or enriched.  Finally a DCBH forms in the core. This was questioned by \citet{2013MNRAS.434.2600F} who claimed that, if a more realistic equation-of-state is adopted, then the disc may cool quickly and the accretion rate drops, leading to a black hole with final mass $\lsim100~M_\odot$. However, a more advanced model has already been investigated by  \citet{2015ApJ...810...51M} and it is found that it is still feasible to form a DCBH via halo merger.  In this paper we only investigate the DCBH formation triggered by external radiation field.

The layout of this paper is as follows. In Section \ref{methods} we introduce the one-zone model and the DCBH SEDs.  In Section \ref{results} we present the results, including tests of our code for a pure blackbody spectrum and for a normal star-forming galaxy spectrum; the critical field intensities for DCBH SEDs and their variance when the XRB is considered. We give the conclusions and discussions in Section \ref{conclusions}. 
In Appendix \ref{optical_depth} we compare the optical depth of the DCBH-forming gas cloud to the LW radiation, to the H$^-$ detachment radiation and to the H$_2^+$ dissociation radiation. In Appendix \ref{window} we present discussions on enhanced DCBH formation probability. In Appendix \ref{reactions} we list the chemical reactions (Table \ref{tb_reactions}) and heating/cooling functions (Tables \ref{tb_heatings} \& \ref{tb_coolings}) included in our chemistry network. Throughout the paper, we use the {\tt Planck} cosmology parameters \citep{2016A&A...594A..13P}: $\Omega_m$=0.308, $\Omega_\Lambda$=0.692, $h$=0.6781, $\Omega_b$=0.0484, $n=0.9677$ and $\sigma_8$=0.8149. The transfer function is from \cite{1998ApJ...496..605E}.

\section{methods}\label{methods}

\subsection{The one-zone model}

The one-zone model developed here describes the evolution of a gas cloud that eventually collapses 
to form a Pop III star, a galaxy or a BH. 
Similar approaches have been successfully applied to a variety of problems  
(see e.g. \citealt{2000ApJ...534..809O, 2001ApJ...546..635O,2005ApJ...626..627O,2014MNRAS.445..544S,2014MNRAS.445..107V,    2015MNRAS.450.4350I,2007MNRAS.382..229S}). 

\subsubsection{Halo build up}
The formation of a dark matter halo is described by the ``top-hat'' spherical collapse scenario, in which the dark matter density $\rho_{\rm d}$ evolves as
\begin{equation}
\rho_{\rm d}=\begin{cases}
       \dfrac{9\pi^2}{2}  \left(  \dfrac{1+z_{\rm ta}}{1-\cos\theta}    \right)^3   \rho_c(\Omega_m-\Omega_b)    &  z > z_{\rm halt} \\
\dfrac{9\pi^2}{2}  (1+z_{\rm ta})^3 \rho_c(\Omega_m-\Omega_b) & z \le z_{\rm halt}, 
\label{eq_rho}
\end{cases}
\end{equation}
where $\rho_c$ is the Universe critical density and $z_{\rm ta}$ is the turn-around redshift; $\theta$ is linked to redshift via 
\begin{equation}
\theta- \sin\theta=\pi\left( \frac{1+z_{\rm ta} }{1+z}\right)^{3/2},
\end{equation}
and by halting the density evolution  at the redshift corresponding to $\theta_{\rm halt}=3\pi/2$, we force the dark matter density not to exceed the final virialization density\footnote{In the literature, in the top-hat spherical collapse scenario, usually the ``virialization redshift" refers to the ``collapse redshift", i.e. the redshift corresponding to $\theta_{\rm vir}=2\pi$ when the density approaches infinity. However this singularity does not happen in practice, since the gas collapse gets halted well before it.}

\begin{figure}
\centering{
\includegraphics[scale=0.4]{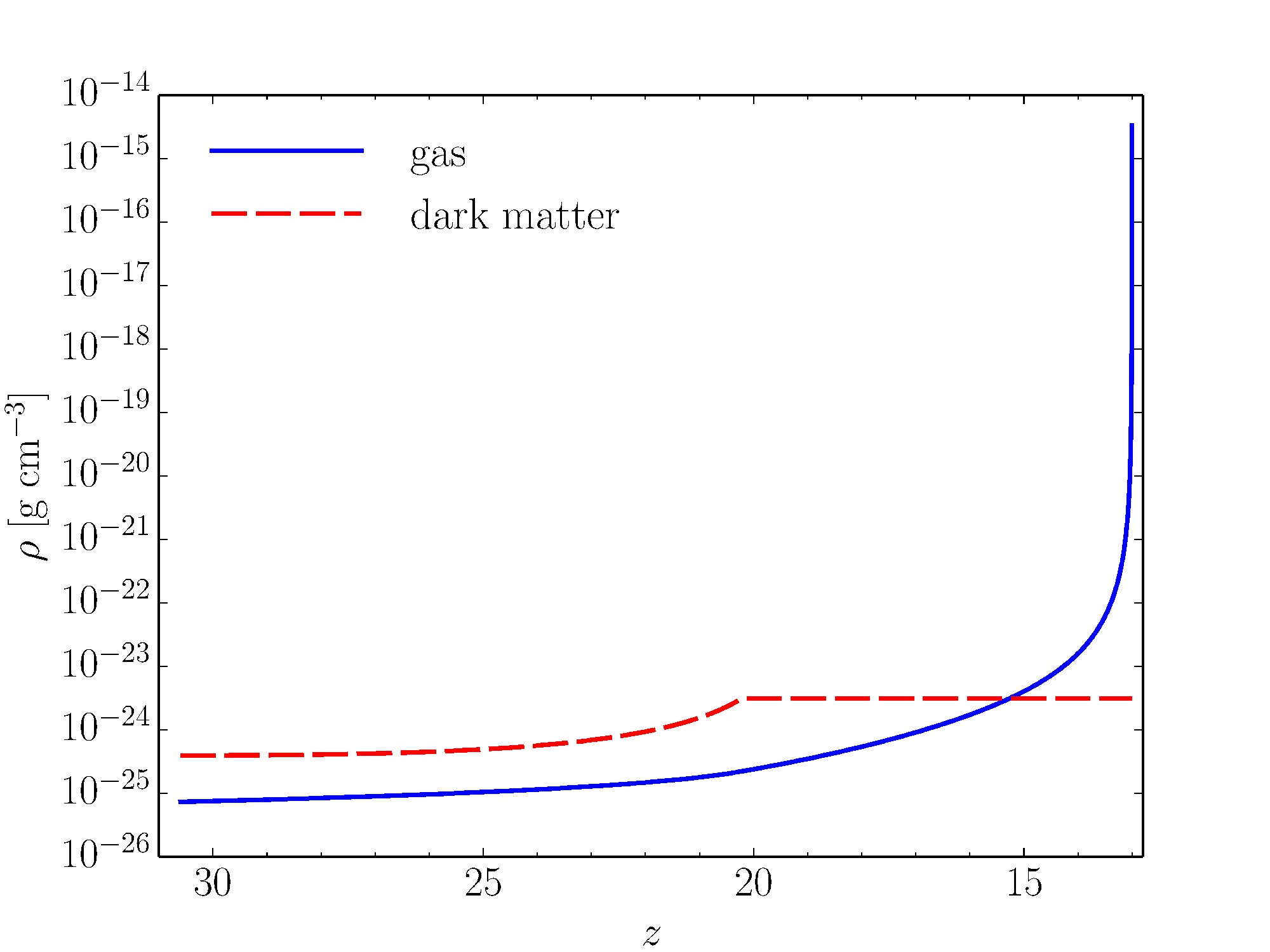}
\caption{The density evolution of dark matter and collapsing gas cloud in a dark matter halo with turn-around redshift 30.6.}
\label{fig_rho}
}
\end{figure}

The dark matter halo provides the gravity that causes its gas content to collapse. Ignoring the pressure, the gas density $\rho_{\rm g}$ evolves as
\begin{equation}
\frac{d\rho_{\rm g}}{dt}=\frac{\rho_{\rm g}}{t_{\rm ff}},
\label{eq_drho_gas} 
\end{equation}
where the free-fall time scale
\begin{equation}
t_{\rm ff} = \sqrt{\frac{3\pi}{32 G(\rho_{\rm d}+\rho_{\rm g})}}.
\end{equation}
Throughout this paper we investigate a halo with $z_{\rm ta} = 30.6$; its gas eventually collapses to form a galaxy or DCBH at $z\sim13$. In Fig. \ref{fig_rho} we plot the evolution of the dark matter and gas density, where the initial gas density is set to be $\Omega_b/(\Omega_m-\Omega_b)\rho_{\rm d}(z_{\rm ta})$. This dark matter collapse gets halted at $z\sim20.5$ and the gas density exceeds the dark matter density at $z\sim15$. 

\subsubsection{Temperature evolution} 

The temperature of the cloud is followed by solving the energy equation 
\begin{equation}
\frac{1}{T}\frac{dT}{dt}=(\gamma-1) \frac{1}{\rho_{\rm g}  }\frac{d\rho_{\rm g}}{dt}+\frac{1}{\mu}\frac{d \mu}{d t}+(\gamma-1)\frac{{\mathcal H}-\Lambda} {nk_{\rm B}T},
\label{eq_dT}
\end{equation}
where $\gamma=5/3$ is the adiabatic index for the monatomic gas, $k_{\rm B}$ is the Boltzmann constant, $\mu$ is the molecular weight and $n=\rho_{\rm g}/\mu$ is the sum of the number density of all the eight species; $\mathcal H$ and $\Lambda$ are the heating and cooling rates per unit volume, 
respectively.

Even in the pre-reionization epoch, X-ray photons can propagate over long distances and heat the gas with a rate
\begin{equation}
{\mathcal H}_{\rm X} =  f_{\mathcal H} {\mathcal E},  
\end{equation}
where ${\mathcal E}$ is the energy (per unit time and volume) of primary electrons due to X-ray ionization, while $f_{\mathcal H}$ is the fraction of the primary energy deposited as thermal energy
and we use the value by \citet{2008MNRAS.387L...8V}  (see \citealt{1985ApJ...298..268S} as well). 
The primary electron energy (here we ignore the photoionization of He$^{+}$)
\begin{align}
{\mathcal E}= \int_{\rm 0.1~keV} dE\frac{4\pi J_X(E)e^{-\tau_{\rm cl}(E)} }{E}  \sum_{i=\rm H,He}(E-E_i)\sigma_i(E)n_i
\end{align}
where $J_X$ is the specific intensity of the X-ray background (see Sec. \ref{XRB}), 
$E_{\rm H} = 13.6$~eV and $E_{\rm He} = 24.6$~eV are the ionization potentials for H and He, respectively.
The photo-electric cross-section, $\sigma_{\rm H}(E)$ and $\sigma_{\rm He}(E)$,  are from \citet{1992ApJ...400..699B}.
The optical depth of the gas cloud is given as 
\begin{equation}
\tau_{\rm cl}(E)= \sum_{i=\rm H,He}\sigma_i(E) n_i\lambda_J/2,
\end{equation}
by using the cloud Jeans length $\lambda_J$.

In addition to the X-ray heating, we also take into account the heating associated with chemical reactions listed in Table \ref{tb_heatings} in Appendix \ref{reactions}. However, throughout this paper we are concerned with DCBH formation in the pre-reionization epoch, when the neutral IGM is opaque to UV photons, we hence ignore the ionization and heating by them.

For the cooling rate $\Lambda$, we include the radiative cooling by H$_2$, H, He, and He$^+$, 
and cooling associated with chemical reactions. 
The cooling functions for individual processes are also listed in Table \ref{tb_coolings} in Appendix \ref{reactions}. 

We stop our calculations at $n\sim10^9$ cm$^{-3}$, because our treatment of H$_2$ cooling is only valid for number densities $\lsim 10^9$ cm$^{-3}$ 
when the gas is optically thin to $\rm H_2$ lines (e.g. \citealt{2006ApJ...652....6Y}). 
However, according to some previous works (e.g. \citealt{2001ApJ...546..635O}), which 
investigated the subsequent evolution up to density $\gg 10^9$ cm$^{-3}$, we can conclude 
that, as long as the H$_2$ formation have been suppressed in a cloud before density reaches $\sim 10^4$ cm$^{-3}$,  
H$_2$ is collisionlly dissociated in higher density and the cloud continues to collapse only via the atomic cooling.
This will result in the formation of a supermassive star at the center, which will then collapses to a BH \citep{2016ApJ...830L..34U}.  

\subsection{Chemical Reactions}

We consider the evolution of eight chemical species H$^{-}$, H, H$^{+}$, H$_2$, H$_2^{+}$, He, He$^{+}$ and $e^{-}$ (we ignore the $\rm He^{++}$ since it plays a negligible role)
by using the reaction rates listed in Table \ref{tb_reactions} in Appendix \ref{reactions}. 
The reaction rates are collected from \citet{1983ApJ...271..632P,  2000ApJ...534..809O, 2008MNRAS.388.1627G,  2010MNRAS.402.1249S} 
and the original references are found in them. 
In a cloud irradiated by an external radiation field, 
H$_2$ formation is suppressed via the following three reactions: 
H$_2$ photo-dissociation
\begin{equation}
{\rm H}_2 + \gamma \xrightarrow{k_{\rm H_2,dis}} 2{\rm H},
\end{equation}
H$^-$ photo-detachment
\begin{equation}
{\rm H}^{-} + \gamma \xrightarrow{k_{\rm H^-, det}}  {\rm H}+e^{-}
\end{equation}
and H$_2^+$ photo-dissociation
\begin{equation}
{\rm H_2}^+ + \gamma \xrightarrow{k_{\rm H_2^+, dis}} {\rm H} + {\rm H}^+
\end{equation}
respectively, where $k_{\rm H_2,dis}, k_{\rm H^-, det}$ and $k_{\rm H_2^+, dis}$ are the corresponding reaction rates 
(i.e., $k_{22}, k_{23}$, and $k_{24}$ in Table \ref{tb_reactions}).
The photo-dissociation and photo-detachement rates of H$_2$, H$^-$ and H$_2^+$ are described below 
in Sec. \ref{H2photo}, \ref{H-photo} and \ref{H2+photo} respectively.
We consider the photo-ionization only by the XRB (Sec. \ref{Xphoto}), 
and ignore all reactions that need ionizing photons other than X-rays.

\subsubsection{Photo-dissociation of $\rm H_2$}\label{H2photo} 
Photons in the narrow LW band can dissociate the $\rm H_2$ through the two-step Solomon process (e.g., \citealt{1967ApJ...149L..29S}). 
The ground electronic state H$_2$ $X(v''=0,J'')$ are excited to the state $B(v',J')$ by absorbing the photons in the Lyman band, or to the state 
$C(v',J')$ by absorbing the photons in the Werner band. 
A fraction of these excited molecules decays to high ($v''\geq 14$) vibrational levels and then dissociates to separate atoms. 
The photo-dissociation rate is then  \citep{1997NewA....2..181A}
\begin{equation}
k_{\rm H_2,dis}=\sum \frac{\pi e^2}{m_{\rm e} c} p_{i'} f_{i'} \int \frac{4\pi J(\nu)}{h \nu}\phi_{i'}(\nu-\nu_{i'})d\nu,
\end{equation} 
where $J(\nu)$ is the specific intensity of the external radiation field at frequency $\nu$, 
$p_{i'}$ the probability of the transition from the ground state to $B(\nu',J')$ or $C(\nu',J')$ \citep{1989A&AS...79..313A}, 
$f_{i'}$ the dissociation fraction \citep{1992A&A...253..525A}, $\nu_{i'}$ the frequency of the transition emission line, 
which has normalized profile $\phi_{i'}(\nu-\nu_{i'})$, and  $\pi e^2/(m_{\rm e}c)=2.65\times10^{-2}$ cm$^2$. 
We assume that the ground-state hydrogen molecules are para-H$_2$ with $J''=0$ \citep{1997NewA....2..181A}. 
In case the line profile is a Dirac $\delta$ function, the above equation reduces to 
\begin{equation}
k_{\rm H_2,dis}\approx  \sum\frac{\pi e^2}{m_{\rm e} c} p_{i'} f_{i'}  \frac{4\pi J(\nu_{i'})}{h\nu_{i'}}.
\label{eq_k_H2_diss}
\end{equation}
The sum is performed for all possible transitions with transition energy smaller than 13.6 eV.
 
For a blackbody radiation field with effective temperature $10^4$ K (``BB") and with mean specific intensity 
$J_{\rm LW}\times 10^{-21}$  erg s$^{-1}$cm$^{-2}$Hz$^{-1}$ sr$^{-1}$ in the LW band\footnote{Note that in the literature sometimes the radiation field strength is represented by different quantities. For example,  in \citet{1997NewA....2..181A} the specific intensity at 12.87 eV is used, while \citet{2010MNRAS.402.1249S} uses the specific intensity at 13.6 eV. We use the \textit{mean} LW specific intensity instead of the specific intensity at a given energy because our DCBH SEDs are not flat in the LW band and because of the presence of several emission lines in this band.}, 
we have
\begin{equation}
k_{\rm H_2,dis}^{\rm BB} =7.74\times10^{-13} J_{\rm LW} f_{\rm sh}~[{\rm s}^{-1}],
\label{eq_k22_BB}
\end{equation}
where the self-shielding effect is included by the self-shielding parameter \citep{2011MNRAS.418..838W}
\begin{equation}
f_{\rm sh}=\frac{0.965}{(1+x/b_5)^{1.1}}+\frac{0.035}{(1+x)^{0.5}}{\rm exp}\left[  -8.5\times10^{-4}(1+x)^{0.5} \right],
\end{equation}
in which $x=N_{\rm H_2}/5\times10^{14}~{\rm cm}^{-2}$ is the H$_2$ column number density $N_{\rm H_2}$ in units $5\times10^{14}$ cm$^{-2}$; and 
$b_5$ is the Doppler broadening parameter in units $10^5$ cm s$^{-1}$
\begin{equation}
b_5=\frac{  \sqrt{2k_{\rm B}T/m_{\rm H_2}}   }{10^5~{\rm cm~s}^{-1}} ;
\end{equation}
and $m_{\rm H_2}$ is the hydrogen molecule mass. We ignore the micro-turbulent  velocity, therefore the self-shielding effect is possibly overestimated here \citep{2001MNRAS.321..385G}.
The H$_2$ column density is given by $N_{\rm H_2}= n_{\rm H_2}\lambda_J/2$, where $\lambda_J$ is the Jeans length of the gas cloud.

For a star-forming galaxy spectrum  (``GAL") with mean LW specific intensity $J_{\rm LW}$,
\begin{equation}
k_{\rm H_2,dis}^{\rm GAL} = 1.23\times10^{-12} J_{\rm LW}f_{\rm sh} ~[{\rm s}^{-1}].
\label{eq_k22_GAL}
\end{equation}
Here, we take the star-forming galaxy spectrum from {\bf \tt STARBURST99}\footnote{http://www.stsci.edu/science/starburst99/docs/default.htm} 
(\citealt{1999ApJS..123....3L,2005ApJ...621..695V,2010ApJS..189..309L}) and adopt the continuous star formation mode, 
Salpeter initial mass function (IMF) with mass range of 0.1 - 100 $M_\odot$, metallicity of $0.02~Z_\odot$ and the age of 100 Myr.

For a Compton-thick DCBH, most photons in the UV to soft X-ray bands are absorbed by the thick envelope enclosing the accretion disc, then re-emitted at the energies below 13.6 eV in the form of Ly$\alpha$ emission, two-photon emission, free-free and free-bound emission, respectively. Therefore its SED has two main components: the remaining unabsorbed photons (hard X-rays and the $<$13.6 eV part of the multi-color blackbody radiation) and the re-processed photons. The relative weight of these two components depends on the column number density of the Compton-thick matter.

\begin{figure}
\centering{
\includegraphics[scale=0.4]{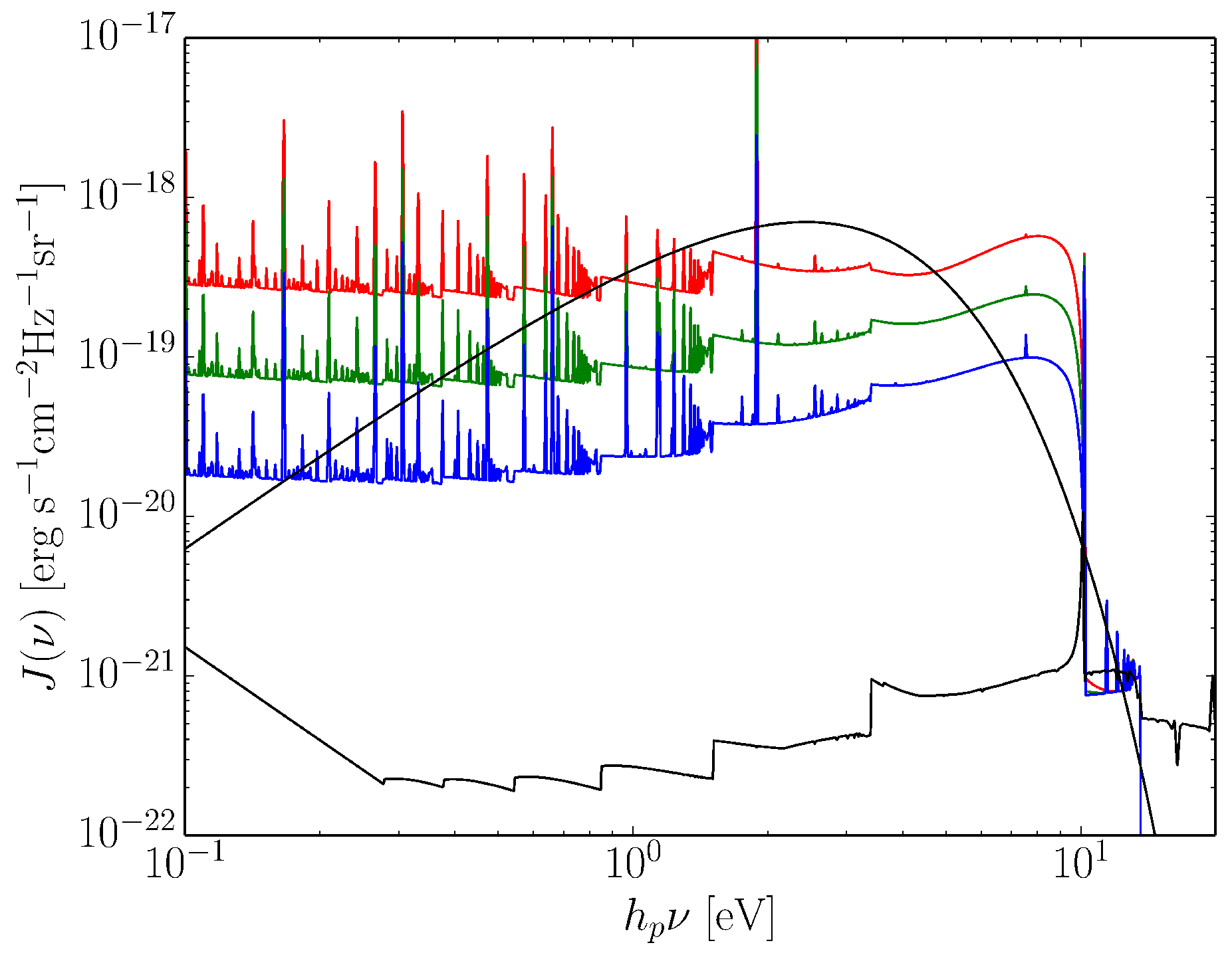}
\caption{The DCBH SEDs (from top to bottom, the three curves with emission lines correspond to BH1, BH2 and BH3 respectively), compared with a blackbody spectrum with effective temperature $10^4$ K (smooth curves) and a star-forming galaxy spectrum (bottom curve with sawteeth).
}
\label{fig_SED_DCBH}
}
\end{figure}

Using numerical simulations \citet{2015MNRAS.454.3771P} have carefully investigated the radiation from DCBHs. They obtained the SEDs for different $N_{\rm H}$ values. In Fig. \ref{fig_SED_DCBH} we show the SEDs of Compton-thick DCBH with $N_{\rm H} = 1.3 \times10^{25}$ cm$^{-2}$ (``BH1"), $8.0\times10^{24}$ cm$^{-2}$ (``BH2") and $5.0\times10^{24}$ cm$^{-2}$ (``BH3") respectively, at the time 10 Myr after the accretion starts. They are for the standard disk, LDP density profiles (the density profile after a BH has already formed at the center) with different normalizations, see \citet{2015MNRAS.454.3771P} for details.
We only plot the energy range relevant to the H$_2$ chemistry, ignoring the $>$13.6 eV part. For all the three SEDs the two-photon emission is the dominant radiation from $\sim3$ eV to $\sim$11.2 eV, below the $\sim3$ eV the free-bound emission dominates over others. The higher the $N_{\rm H}$, the more energy is re-processed to the $<$13.6 eV part, and such SED is expected to suppress the H$_2$ formation more efficiently. In case the envelope is rather thick, i.e. $N_{\rm H}\gsim10^{25}$ cm$^{-2}$, Ly$\alpha$ emission is barely seen as the corresponding photons are well-trapped in the thick envelope and, finally most of them escape from the envelope as two-photon emission. 
 
The SED obtained in  \citet{2015MNRAS.454.3771P} evolves as the accretion progresses. However, for the BH mass investigated in this paper the SED is always stationary when the accretion time is smaller than $\sim100$~Myr, only the normalization changes with the increasing BH mass. In our work we always use the SEDs plotted in Fig. \ref{fig_SED_DCBH} as the spectrum of the external radiation field, leaving the normalization as a free parameter representing the different distances from the collapsing gas cloud to the nearby emitting DCBH.

For comparison  the BB and GAL spectra are also shown in Fig. \ref{fig_SED_DCBH}. All SEDs there have  $J_{\rm LW} = 1.0$. 
Substituting the DCBH SEDs into Eq. (\ref{eq_k_H2_diss}), we find all the three DCBH models have almost the same H$_2$ photo-dissociation rate
\begin{align}
k_{\rm H_2,dis}^{\rm BH} &=1.30\times10^{-12} J_{\rm LW} f_{\rm sh}~[{\rm s}^{-1}].
\label{eq_k22_BH}
\end{align}

\subsubsection{Photo-detachment of $\rm H^-$} \label{H-photo} 

The H$^-$ could be detached by photons in the wide energy range of 0.755 - 13.6 eV. The reaction rate is obtained through integration,
\begin{equation}
k_{\rm H^-, det} = \int_{\nu_{0.755}}^{\nu_{13.6}} \frac{4\pi J(\nu)}{h\nu} \sigma_{\rm H^-,det}(\nu)d\nu,
\label{eq_k23}
\end{equation}
where the cross-section is \citep{1997NewA....2..181A}
\begin{equation}
\sigma_{\rm H^-,det}(\nu)=7.928\times10^5(\nu-\nu_{0.755})^{1.5}/\nu^3~[{\rm cm^2}];
\end{equation}
$\nu_{0.755}$ is the frequency (in Hz) of a photon with energy 0.755 eV and $\nu_{13.6}$ with 13.6 eV. For the BB and GAL radiation fields with mean LW specific intensity $J_{\rm LW}$, we obtain\footnote{Here the mean LW specific intensity, $J_{\rm LW}$, is only used to denote the normalization of the SED, actually all photons with energy $\gsim0.755$ eV detach the $\rm H^-$.} 
\begin{equation}
k_{\rm H^-, det}^{\rm BB} = 5.60\times10^{-8} J_{\rm LW}~[{\rm s^{-1}}],
\label{eq_k23_BB}
\end{equation}
and 
\begin{equation}
k_{\rm H^-, det}^{\rm GAL} = 5.43\times10^{-11} J_{\rm LW}~[{\rm s^{-1}}].
\label{eq_k23_GAL}
\end{equation}
For DCBH SEDs,
\begin{align}
k_{\rm H^-, det}^{\rm BH1}&=5.27\times10^{-8}J_{\rm LW}~[{\rm s^{-1}}] \nonumber \\
k_{\rm H^-, det}^{\rm BH2}&=1.87\times10^{-8}J_{\rm LW}~[{\rm s^{-1}}] \nonumber \\
k_{\rm H^-, det}^{\rm BH3}&=6.29\times10^{-9}J_{\rm LW}~[{\rm s^{-1}}].
\label{eq_k23_BH}
\end{align}
Apparently, $k_{\rm H^-, det}^{\rm BH1}\approx k_{\rm H^-, det}^{\rm BB} \gg k_{\rm H^-, det}^{\rm GAL} $, and since this is the most relevant reaction suppressing the $\rm H_2$ formation, we therefore expect the critical specific intensity, $J_{\rm LW}^{\rm crit}$, for a BH1 radiation field, to be comparable with a BB radiation field but it is much smaller than a GAL radiation field.

\subsubsection{Photo-dissociation of $\rm H_2^+$}\label{H2+photo} 

Here we present the photo-dissociation rate of the H$_2^+$, a more careful investigation of the role played by H$_2^+$ channel in H$_2$ formation can be found in \citet{2016MNRAS.456..270S}, who pointed out that this channel is even more important than the H$^-$ channel in the softer radiation field.
$\rm H_2^+$ is dissociated by photons above 2.65 eV. The reaction has a cross-section  \citep{1987ApJ...318...32S} given by
\begin{align}
&{\rm log}(\sigma_{\rm H_2^+})=  
\begin{cases}
 -40.97+6.03E_\gamma-0.504E_\gamma^2+1.387\times10^{-2}E_\gamma^3 \nonumber \\
  \hfill (2.65 <E_\gamma <11.27 )\nonumber  \\
 -30.26+2.79E_\gamma-0.184E_\gamma^2+3.535\times10^{-3}E_\gamma^3  \nonumber \\
\hfill (11.27<E_\gamma < 21) 
\end{cases}   
\end{align}
where $E_\gamma$ and the cross-section are in units eV and cm$^2$ respectively. Similar to Eq. (\ref{eq_k23}), we get the reaction rate for the BB and GAL radiation fields
\begin{equation}
k_{\rm H_2^+, dis}^{\rm BB} =4.30\times10^{-11}J_{\rm LW}~[{\rm s^{-1}}], 
\label{eq_k24_BB}
\end{equation}
and 
\begin{equation}
k_{\rm H_2^+, dis}^{\rm GAL} = 6.25\times10^{-12} J_{\rm LW}~[{\rm s^{-1}}].
\label{eq_k24_GAL}
\end{equation}
For DCBH SEDs
\begin{align}
k_{\rm H_2^+, dis}^{\rm BH1} &=8.61\times10^{-10}J_{\rm LW}~[{\rm s^{-1}}] \nonumber \\ 
k_{\rm H_2^+, dis}^{\rm BH2} &=3.63\times10^{-10}J_{\rm LW}~[{\rm s^{-1}}] \nonumber \\
k_{\rm H_2^+, dis}^{\rm BH3} &=1.46\times10^{-10}J_{\rm LW}~[{\rm s^{-1}}].
\label{eq_k24_BH}
\end{align}
We have confirmed that the gas cloud is always transparent to both H$^-$ detachment radiation and H$_2^+$ dissociation 
radiation and neglect their shielding effect, see a check in Appendix \ref{optical_depth}.

\subsubsection{Photo-ionization of {\rm H} and {\rm He} by X-rays} \label{Xphoto}

The total ionization rate of hydrogen atoms, including photoionization and secondary ionizations,  is  
\begin{equation}
{\mathcal I}_{\rm H}=   \int dE\frac{4\pi J(E)e^{-\tau_{\rm cl}(E)} }{E}    \sigma_{\rm H}(E)n_{\rm H}+   \frac{ f_{ {\mathcal I},\rm H}   {\mathcal E} }{E_{\rm H}},
\end{equation}
and of the helium atoms is
\begin{equation}
{\mathcal I}_{\rm He}=   \int dE\frac{4\pi J(E)e^{-\tau_{\rm cl}(E)} }{E}    \sigma_{\rm He}(E)n_{\rm He}+   \frac{ f_{ {\mathcal I},\rm He}   {\mathcal E} }{E_{\rm He}},
\end{equation}
where $  f_{ {\mathcal I},\rm H} $ and $  f_{ {\mathcal I},\rm He} $  are fractions of primary energy deposited into hydrogen secondary ionization and helium secondary ionization respectively and are from \citet{2008MNRAS.387L...8V}. 
 
\subsubsection{Numerical Approach}
The initial temperature is set to be the adiabatic temperature of the IGM at $z_{\rm ta}$ \citep{2001PhR...349..125B}. The initial H$_2$ fraction, i.e. the H$_2$ number density to H and H$^+$ number density ratio,  is set to be $10^{-6}$ \citep{1984ApJ...280..465L,2006MNRAS.372.1175H}; the initial H$^+$ fraction and He$^+$ fraction are all set to be $10^{-4}$, and the initial electron fraction is their sum; fractions of other species are set to be zero.
 
We solve the differential equations of the chemistry network together with Eq. (\ref{eq_drho_gas}) and Eq. (\ref{eq_dT}) starting from $z_{\rm ta}$, to obtain the evolution of gas density, temperature and chemical abundances. To guarantee the precision we force the time-step to be the 0.1\% of the minimum of all $y_i/(dy_i/dt)$, where $dy_i/dt$ is the derivative of the $i$th differential equation excluding $d$H$^-/dt$ and $d$H$_2^+/dt$ (see below). 
The reaction rates for the intermediary species, H$^-$ and H$_2^+$, are much larger than for the other species. 
Hence, to reduce the computational time, we always use equilibrium abundances for H$^-$ and H$_2^+$ obtained by iteratively solving the following simultaneous equations:
\begin{align}
n^{\rm eq}_{\rm H^-}&=(k_7n_{\rm H}n_e+k_{12}n_{\rm H_2}n_e)/(k_{8}n_{\rm H}+k_{10}n_{\rm H_2^+}^{\rm eq}+k_{11}n_{\rm H^+}\nonumber \\
&+k_{17}n_{\rm e}+k_{18}n_{\rm H} +k_{19}n_{\rm H^+}+k_{\rm H^-, det}), \nonumber \\
n^{\rm eq}_{\rm H_2^+} &=(k_5n_{\rm H}n_{\rm H^+} + k_{15}n_{\rm H_2}n_{\rm H^+}+k_{19}n_{\rm H^-}^{\rm eq}n_{\rm H^+} )/(k_6n_{\rm H} \nonumber \\
&+k_9n_{\rm e}+k_{10}n_{\rm H^-}^{\rm eq}+k_{\rm H_2^+, dis}).
\end{align}
The validity of this treatment has been proven in previous works (e.g. \citealt{1997NewA....2..181A, 1998ApJ...508..141O,2002ApJ...564...23B,2010ApJ...722.1793O}).

\section{Results}
\label{results}
\subsection{The critical intensity for the blackbody and galactic spectra: Tests of the code}\label{test_code}

First, we run the code for a collapsing gas cloud irradiated by the BB and the GAL radiation fields as a test. 
Here we ignore the XRB.  

\begin{figure}
\centering{
\subfigure{\includegraphics[scale=0.4]{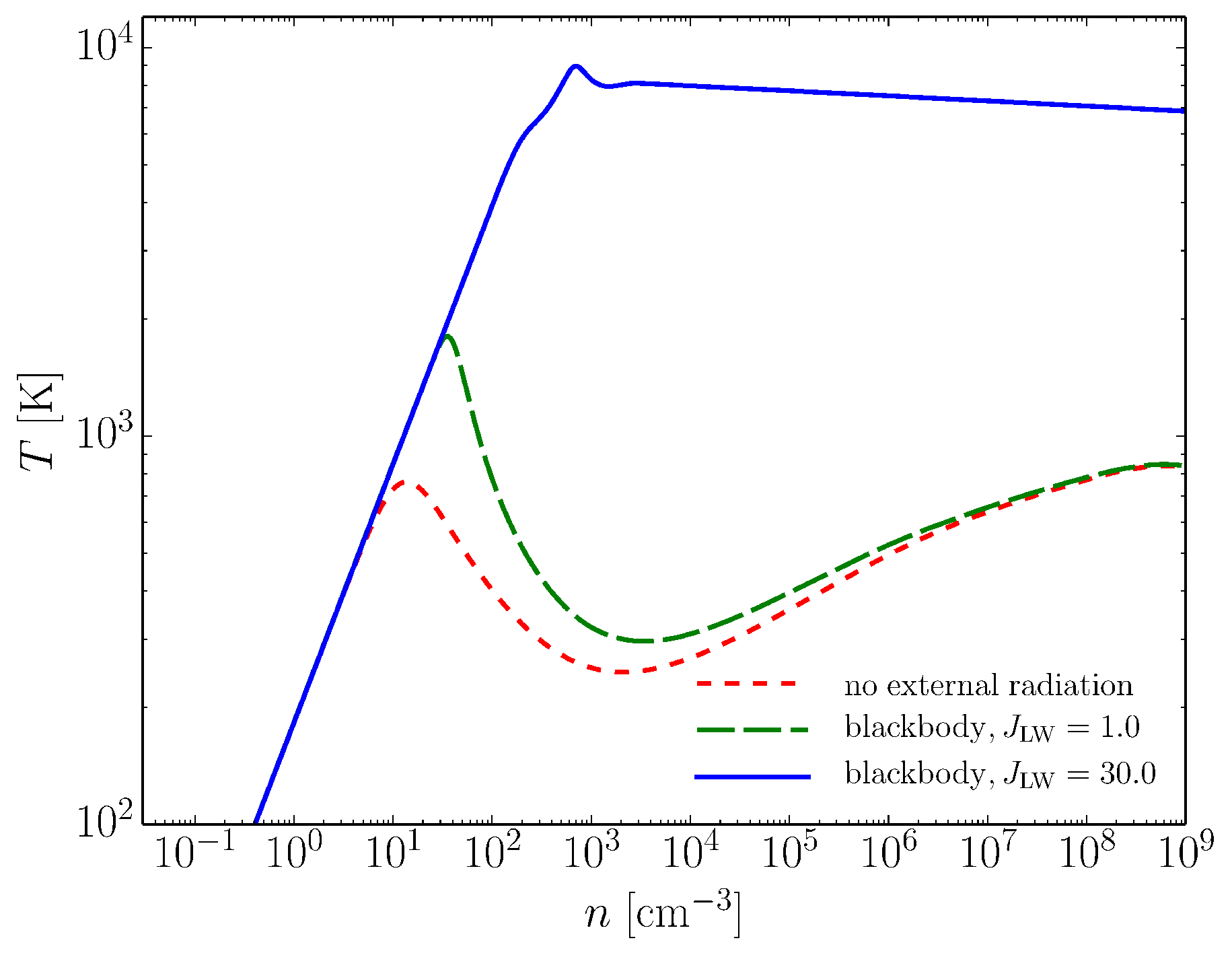}}
\subfigure{\includegraphics[scale=0.4]{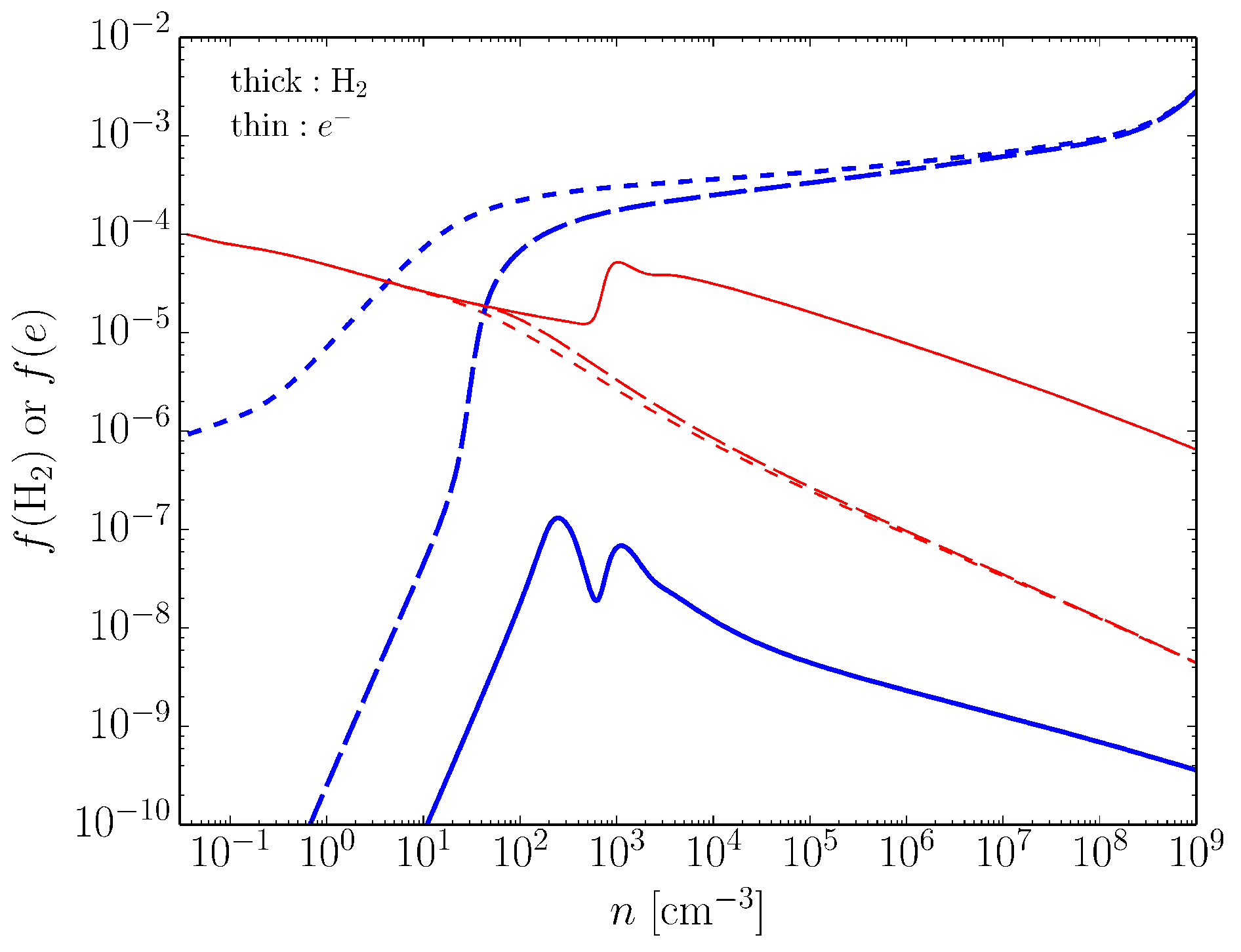}}
\caption{{\it Upper panel:} Temperature evolution as a function of gas density for a gas cloud irradiated by an external BB radiation field
with different $J_{\rm LW}$ values. {\it Bottom:} Evolution of the H$_2$ (thick lines) and $e^-$ (thin lines) fractions for the BB model.}
\label{fig_Tgas_BB1e4K}
}
\end{figure}

For the BB radiation field with different strengths, the temperature, the H$_2$ and $e^-$ fractions are plotted 
in upper and bottom panels of Fig. \ref{fig_Tgas_BB1e4K}, respectively.
In the absence of an external radiation field or when the radiation field is weak (e.g. $J_{\rm LW} = 1$), 
the gas cloud cools by H$_2$ and attains a temperature $\sim 250-300$~K at $n \sim10^3$ cm$^{-3}$, 
where the H$_2$ rotational levels reach the local thermodynamic equilibrium.  
At higher densities, H$_2$ cooling rate saturates and the temperature increases gradually, reaching $\sim 800$~K at $n \sim 10^9$ cm$^{-3}$, 
consistent with previous studies \citep{2001ApJ...546..635O}.  
On the other hand, with a more intense field, i.e., $J_{\rm LW} = 30$, the H$_2$ formation and cooling are suppressed, 
and up to $n\sim10^9$ cm$^{-3}$ the evolution remains quasi-isothermal at $T \gsim 6000$ K set by atomic-cooling mechanism. 

The blackbody spectrum with effective temperature $10^4$ K has $J_{\rm LW}/J_{\nu_{13.6}} = 3.69$. 
Hence $J_{\rm LW}^{\rm crit} = 30$ is translated into $J^{\rm crit}_{\nu_{13.6}} = 8$. 
This is smaller than \citet{2010MNRAS.402.1249S} who found $J^{\rm crit}_{\nu_{13.6}} = 39$. 
\citet{2014MNRAS.443.1979L,2015MNRAS.451.2082G} pointed out that \citet{2010MNRAS.402.1249S} ignored the dissociative tunneling effect term in \citet{1996ApJ...461..265M}, as a result their H$_2$ collisional dissociation rate is underestimated. 
Our critical value is even smaller than \citet{2015MNRAS.451.2082G} who got $J^{\rm crit}_{\nu_{13.6}} \approx17$. 
We have checked that such discrepancy is due to the different initial setups. Using the same initial setups,  and using $N_{\rm H_2}=n_{\rm H_2}\lambda_J$ instead of $N_{\rm H_2}=n_{\rm H_2}\lambda_J/2$, we get $J_{\nu_{13.6}}^{\rm crit}=16$, very close to \citet{2015MNRAS.451.2082G}. We further test the same initial setups as \citet{2011MNRAS.416.2748I}, and consistently get  $J_{\nu_{13.6}}^{\rm crit}=16$. 

In addition to the BB radiation field, we also check the GAL field, and find that when $J_{\rm LW} \gsim 700$, the H$_2$ formation and cooling is suppressed.
 
Regarding a Compton-thin black hole, its SED is composed of two parts with comparable bolometric luminosities \citep{2005MNRAS.362L..50S}: a multi-color black body spectrum that dominates below $\lsim 0.2$ keV, and a power-law spectrum that dominates at $\gsim 0.2$ keV.  The corresponding H$_2$ photo-dissociation rate is $\approx1.26\times10^{-12}J_{\rm LW}f_{\rm sh}~[\rm s^{-1}]$  and the H$^-$ photo-detachment rate is $\approx6.84\times10^{-11}J_{\rm LW}{\rm s^{-1}}$, and less sensitive to the black hole mass. The rates are similar to a GAL field. We therefore suspect that the critical field strength for a Compton-thin black hole is close to the GAL field.  However, in addition to the photons that dissociate the H$_2$ and detach the H$^-$, a Compton-thin black hole simultaneously emits lots of UV and soft X-ray photons. These photons ionize and heat the collapsing gas cloud, and may either destroy it or enhance the H$_2$ formation therein (e.g. \citealt{2016MNRAS.461..111R}), resulting in a rather different critical field strength via complex mechanisms. For this reason we do not investigate such a SED in this test. 

\subsection{The critical intensity by DCBHs: cases without XRB}

\begin{figure}
\centering{
\includegraphics[scale=0.4]{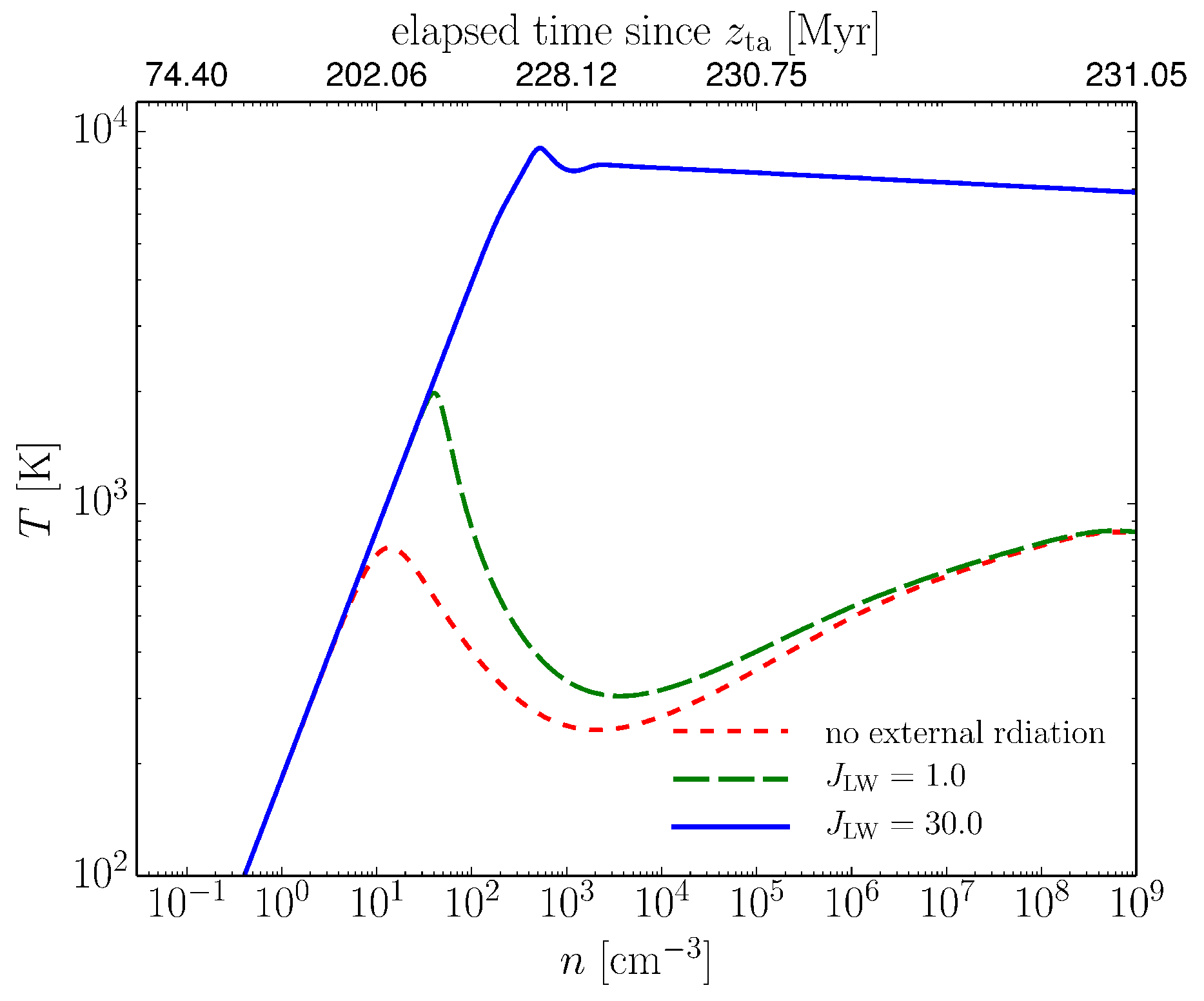}
\caption{Temperature evolution for the gas cloud irradiated by the BH1 radiation field 
with different strengths. The model without external radiation field is also plotted. 
We mark the elapsed time since $z_{\rm ta}$ on the upper x-axis.}
\label{fig_Tgas_DCBH}
}
\end{figure}

In Fig. \ref{fig_Tgas_DCBH}, we show the temperature evolution of a collapsing gas cloud 
irradiated by the BH1 radiation field, for different field strengths, which corresponds to 
different distances to the source DCBH.  
For this model, we find $J_{\rm LW}^{\rm crit}\approx22$ and, as expected, 
it is similar to the BB radiation 
since $k_{\rm H^-, det}^{\rm BB} \approx k_{\rm H^-, det}^{\rm BH1}$. 
For the BH2 and BH3 spectra,
we find $J_{\rm LW}^{\rm crit}\approx35$ and $J_{\rm LW}^{\rm crit}\approx54$, respectively.

In the above calculations the external radiation field is turned on from the initial phase, 
i.e. $z_{\rm ta} = 30.6$, to the final collapse redshift, i.e. $z_{\rm fin} = 13$. 
The time interval between these two epochs is $\approx233$~Myr. 
This might raise some concern as such time span is longer than the typical DCBH accretion time, 
$\sim 100$ Myr. We note, however, that such a long irradiation time is not really necessary. 
For most of the time the gas density and temperature remain around the initial values.  
Later on, the cumulative H$_2$ formed during this early evolutionary stages will be easily 
washed out by external radiation field as long as it can penetrate into the gas cloud. 

\begin{figure}
\centering{
\includegraphics[scale=0.4]{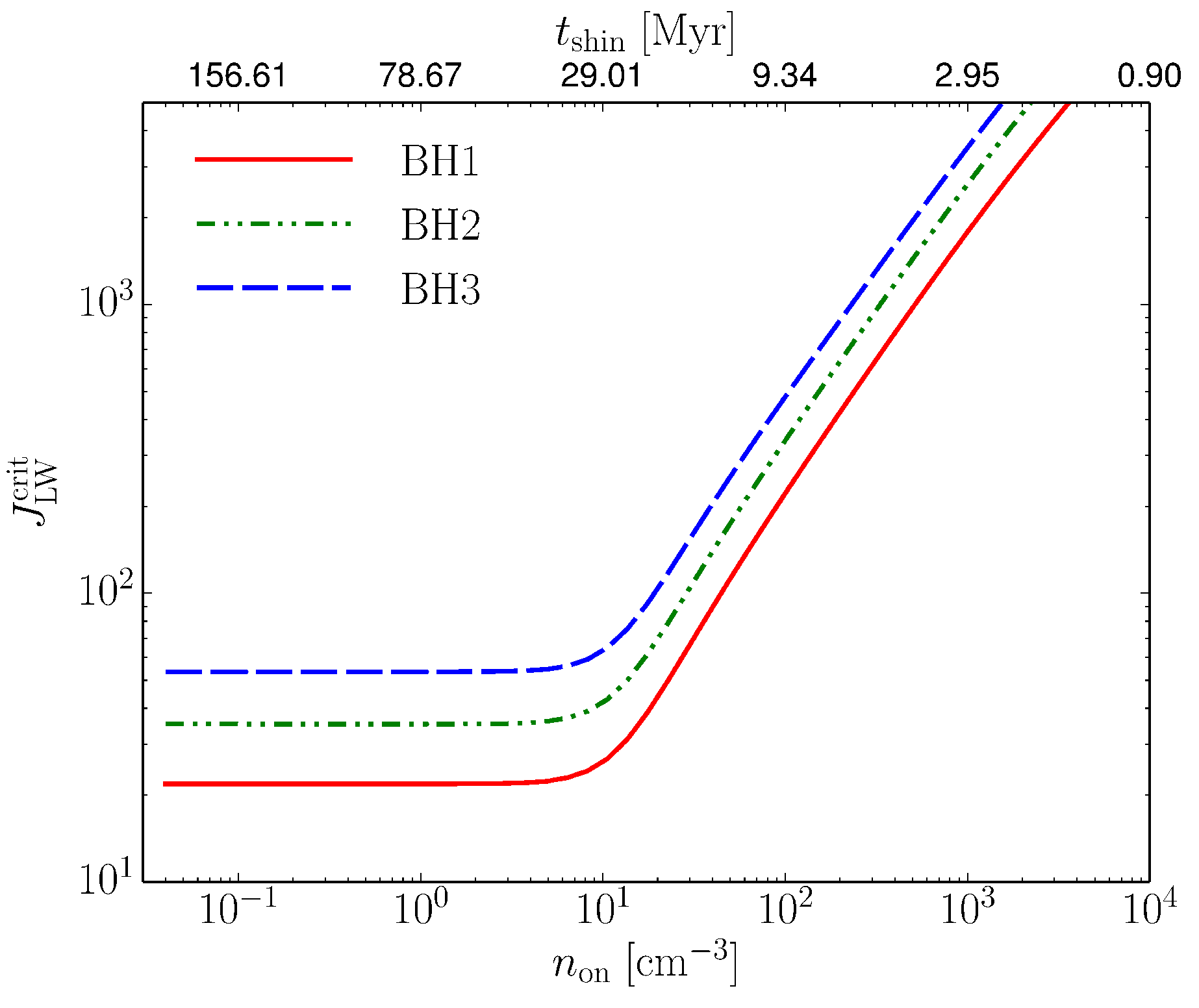}
\caption{The critical intensity $J_{\rm LW}^{\rm crit}$ vs. the density at which the radiation field 
is switched on, $n_{\rm on}$, for different DCBH spectra. The shining time is marked on the upper 
$x$-axis.}
\label{fig_J21_ntot_turnon}
}
\end{figure}

We now examine the minimal requirement for the irradiation time.  
For this purpose, we initially set the external radiation to be zero, and switch it on 
at some gas density $n_{\rm on}$. 
We then repeat the calculations for different $n_{\rm on}$ values
to obtain the dependence of $J_{\rm LW}^{\rm crit}$ on $n_{\rm on}$. 
The result is presented in Fig. \ref{fig_J21_ntot_turnon}. 
The shining time $t_{\rm shin}$, defined as the duration of the irradiation
before the final collapse, is also indicated on the upper $x$-axis. 
For $n_{\rm on} \la 10$ cm$^{-3}$, $J_{\rm LW}^{\rm crit}$ remains constant. 
While for higher $n_{\rm on}$, the increasingly higher $J_{\rm LW}^{\rm crit}$ is needed 
for the direct collapse. 
For example, with the BH1 spectrum, for $n_{\rm on}$ as high as $100$ cm$^{-3}$, 
$J_{\rm LW}^{\rm crit}$ is raised to $\approx230$. 
In this case, the cloud can be irradiated only for $\sim 9$ Myr 
(from $n = 100$ cm$^{-3}$ to the final collapse). 
If the external radiation field is switched on at $n_{\rm on} = 1000$ cm$^{-3}$, the required 
intensity is even higher, $J_{\rm LW}^{\rm crit}\approx1800$. 
In this case external radiation field is only required for $\sim 3$ Myr. 
  
\begin{deluxetable}{ccccc}
\tablecaption{fitting coefficients for $J^{\rm crit}_{\rm LW}$ - $n_{\rm on}$ relation}\label{tab_coef}
\tablehead{\colhead{spectrum} & \colhead{$c_1$} &\colhead{$c_2$} & \colhead{$c_3$} & \colhead{$c_4$}} 
\startdata 
BH1 & 1.34 & 23.3 & 0.83 &$-0.60$ \\
BH2 & 1.55 & 22.3 & 0.82 &$-0.60$ \\
BH3 & 1.73 & 21.8 & 0.80 &$-0.60$ 
\enddata
\label{tab_coef}
\end{deluxetable}

The $J^{\rm crit}_{\rm LW}$ - $n_{\rm on}$ relation shown in Fig. \ref{fig_J21_ntot_turnon}
can be fitted by a formula 
\begin{equation}
{\rm log}(J_{\rm LW}^{\rm crit})=c_1+(1-e^{-n_{\rm on}/c_2})[ c_3 {\rm log}(n_{\rm on})+c_4 ] 
\end{equation}
for each type of the spectrum and the coefficients $c_{1}-c_{4}$ are presented 
in Table \ref{tab_coef}.  
It may be more convenient to know the $t_{\rm shin}$ in some cases, we therefore also fit a formula for  the $J_{\rm LW}^{\rm crit} - t_{\rm shin}$ relation, 
\begin{equation}
{\rm log}(J_{\rm LW}^{\rm crit})=d_1+(1-e^{-d_2/t_{\rm shin}})[ d_3\times {\rm log}(t_{\rm shi})+d_4],
\end{equation}
parameters are listed in Tab. \ref{tab_coef2}.

\begin{deluxetable}{ccccc}
\tablecaption{fitting coefficients for $J^{\rm crit}_{\rm LW}$ - $t_{\rm shin}$ relation}\label{tab_coef2}
\tablehead{\colhead{spectrum} & \colhead{$d_1$} &\colhead{$d_2$} & \colhead{$d_3$} & \colhead{$d_4$}} 
\startdata 
BH1 & 1.34 & 14.6 & $-1.49$ &2.65 \\
BH2 & 1.55 & 14.5 & $-1.45$ &2.59 \\
BH3 & 1.73 & 14.5 & $-1.43$ &2.53 
\enddata
\label{tab_coef2}
\end{deluxetable}

\subsection{The impact of an XRB}\label{XRB}
The XRB ionizes the cloud, producing more free electrons and promoting $\rm H_2$ formation. 
As such, it works as a negative feedback on DCBH formation and boost the critical intensity 
\citep{2011MNRAS.416.2748I}. 
Here, we study to what extent the critical intensity is modified 
by the presence of a XRB. 
We consider two types of the X-ray sources: (i) accreting DCBHs and (ii) first galaxies. 

\subsubsection{X-rays from DCBHs}\label{XRB_DCBH}
Since a DCBH is Compton-thick during most of its accretion stage, 
soft X-rays are absorbed in the envelope. 
Only hard X-rays come out from the source, which have negligible impacts on a nearby forming DCBH. 
However, near the end of accretion, with most of the halo gas reservoir having been swallowed 
by the growing DCBH, some soft X-ray photons can leak out of the system \citep{2015MNRAS.454.3771P}.  
Such soft X-rays from DCBHs cumulatively builds up a high-$z$ XRB.

We first derive the XRB spectrum from accreting DCBHs.
Assuming that the DCBH formation rate is proportional to the formation rate 
of atomic-cooling halos with the virial temperature 
in range 10000~K $<T_{\rm vir}<$ 20000~K, 
the DCBH X-ray emissivity is given by
\begin{equation}
e_X^\bullet(E',z)=\frac{1}{4\pi}\int_z     L_X^\bullet(E',\Delta t')   f_\bullet \frac{d{\mathcal N}_{\rm h}}{dz'}dz',
\end{equation}
where ${\mathcal N}_{\rm h}$ is the number density of atomic-cooling halos, 
$f_\bullet$ their fraction harboring DCBHs, 
$\Delta t'$ the time interval between redshifts $z$ and $z'$. 
We take the time-dependent $L_X^\bullet$ of a DCBH with initial mass $2\times10^5~M_\odot$ 
from \citet{2015MNRAS.454.3771P}'s numerical simulations. 
Since we are mainly interested in the XRB spectral shape, its amplitude is left as a free parameter.
To express different XRB levels, 
we vary the parameter $f_\bullet$, 
which is also related to the accretion rate density by 
\begin{equation}
\dt{\rho}_{\bullet}(z)=\int_z   \dt{M}_{\bullet}(\Delta t')   f_\bullet   \frac{d{\mathcal N}_{\rm h}}{dz'}dz',
\end{equation}
where $\dt{M}_\bullet(\Delta t')$ is the accretion rate onto a BH 
at a time interval $\Delta t'$ after its birth and taken from the numerical simulations in \citet{2015MNRAS.454.3771P}\footnote{Strictly speaking, BHs with different initial masses have different growth histories. Although during the growth $L_X^\bullet$ is found to be almost proportional to the BH mass in \citet{2015MNRAS.454.3771P} simulations, the final XRB may depend on both the number fraction of BHs in newly-formed atomic-cooling halos and the initial mass function of BHs.}.
There are large uncertainties on the accretion rate density $\dt{\rho}_\bullet$, because both the occupation fraction of DCBHs in atomic-cooling halos and the BH mass distribution are not well-known. In the following, we therefore consider $\dt{\rho}_\bullet$ as a free parameter.  
The specific intensity of XRB at redshift $z$ is then
\begin{equation}
J_X^\bullet(E,z)= (1+z)^3\int_{z}  e_X^\bullet(E',z')e^{-\tau_{\rm IGM}(E')} \frac{dr_p}{dz'}dz',
\label{eq_JX_DCBH}
\end{equation}
where $E'=E(1+z')/(1+z)$, and $r_p$ is the proper distance. The IGM optical depth for X-ray photons is
\begin{align}
\tau_{\rm IGM}(E') &=\int_{z}^{z'} 
\left[ \sum_{i={\rm H, He}} \bar{n}_{i}(z'')\sigma_{i} (E'')  \right] 
\frac{dr_p}{dz''}dz'',
\end{align}
where$E''=E (1+z'')/(1+z)$ and 
$\bar{n}_{i}$ is the mean number density of species $i$ in the IGM 
(the trace amount metals in the IGM are ignored).

\begin{figure}
\centering{
\includegraphics[scale=0.4]{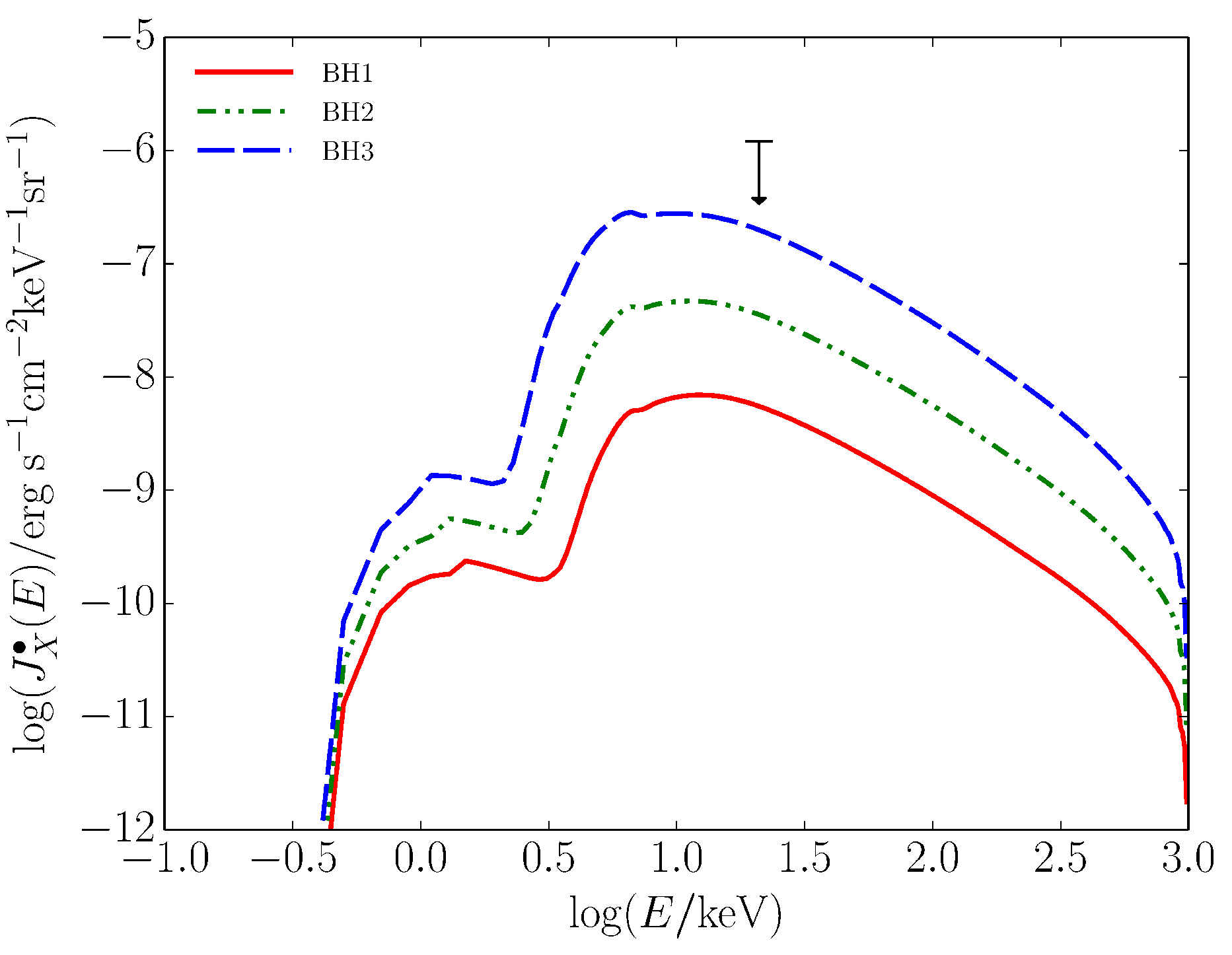}
\caption{
The spectral energy distributions of the XRB from DCBH populations 
with shielding columns $N_{\rm H} = 1.3 \times10^{25}$ cm$^{-2}$ (``BH1''), 
$8.0\times10^{24}$ cm$^{-2}$ (``BH2'') and $5.0\times10^{24}$ cm$^{-2}$ (``BH3'') 
from the bottom to top, respectively, 
assuming the BH accretion rate density 
$\dot{\rho}_\bullet= 1\times10^{-3} M_\odot {\rm yr}^{-1} {\rm Mpc}^{-3}$ at $z = 13$. 
The arrow is the upper limit derived from the present-day observed XRB level at 1.5 keV 
by \citet{2012A&A...548A..87M} }
\label{fig_J_X}
}
\end{figure}

We plot the XRBs at $z=13$ for the three types of source DCBH spectra in Fig. \ref{fig_J_X} 
for $\dt{\rho}_\bullet = 10^{-3}~M_\odot {\rm yr^{-1}Mpc^{-3}}$. 
From the fact that the present-day intensity does not fall below the redshifted intensity from the past, 
i.e., 
\begin{equation}
J_X(E,z=0)\gsim (1+z)^{-3}J_X(E(1+z),z),
\end{equation}
we can put a constraint on the XRB at high redshift
from the present-day XRB from unresolved sources,
$EJ_X(1.5{\rm~keV}, z=0)=2\times10^{-13}$ erg s$^{-1}$cm$^{-2}$deg$^{-2}$ \citep{2012A&A...548A..87M}.
The upper limit thus obtained for XRB at $z=13$, 
$J(21{\rm~keV}, z=13)<1.2\times10^{-6}$ erg s$^{-1}$cm$^{-2}$keV$^{-1}$sr$^{-1}$, 
which is shown in Fig. \ref{fig_J_X} by a downward arrow. 

\begin{figure*}
\centering{
\includegraphics[scale=0.8]{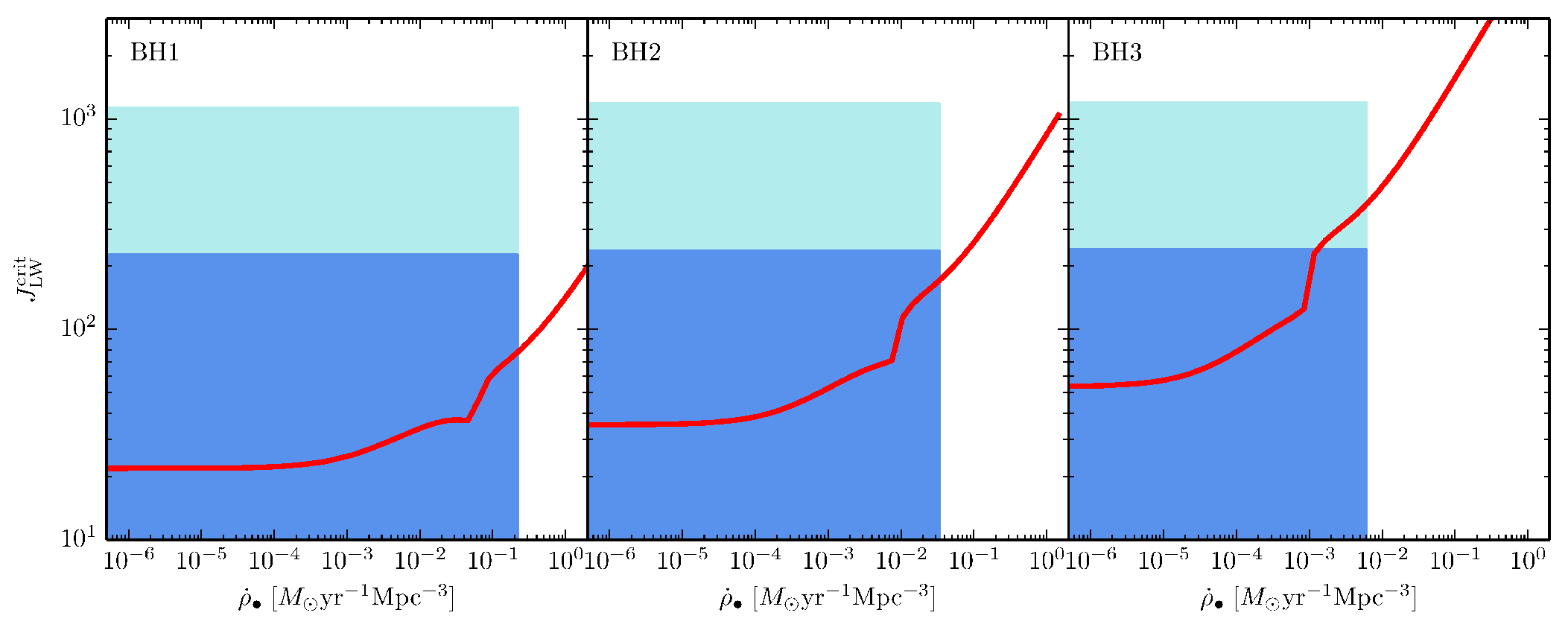}
\caption{$J_{\rm LW}^{\rm crit}$ as a function of $\dot{\rho}_\bullet$ for three DCBH models. 
We fill the regions where the XRB produced by $\dot{\rho}_\bullet$ is below the present-day XRB level constraints, and the required $J_{\rm LW}^{\rm crit}$ can be provided by an emitting DCBH with mass $2\times10^5~M_\odot$ (dark shaded) and $1\times10^6~M_\odot$ (light \& dark shaded). 
}
\label{fig_J21_vs_rho_acc}
}
\end{figure*}

Armed with the XRB spectrum above, 
we next investigate the impact of XRB on the DCBH formation. 
By repeating the temperature evolution calculations for one of 
the three DCBH spectra 
(BH1-3 with $N_{\rm H}=1.3 \times 10^{25}, 8.0 \times10^{24}$, and $5.0 \times 10^{24} {\rm cm^{-2}}$) 
with the XRB of the same $N_{\rm H}$, 
we can derive the critical intensity $J_{\rm LW}^{\rm crit}$ for the DCBH formation 
as a function of $\dt{\rho}_\bullet$.
The result is shown in Fig. \ref{fig_J21_vs_rho_acc} for the three DCBH spectra. 
For the same XRB level (i.e., $\dt{\rho}_\bullet$), 
the $J_{\rm LW}^{\rm crit}$ increases with decreasing $N_{\rm H}$
as in the cases without the XRB.
In the case of BH1, 
$J_{\rm LW}^{\rm crit} $ remains $\approx$ 22 below 
$\dt{\rho}_\bullet \sim 10^{-4}~M_\odot {\rm yr^{-1} Mpc^{-3} }$.
The behavior of $J_{\rm LW}^{\rm crit}$ is similar also for BH2(3) cases:
$J_{\rm LW}^{\rm crit}$ is always $\approx$ 35 (54)
below $\dt{\rho}_\bullet \sim 10^{-5}$($10^{-6}$)$~M_\odot {\rm yr^{-1} Mpc^{-3}}$.
 
Assuming that active DCBHs accrete at the Eddington limited rate\footnote{Note, however, that simulations by \citealt{2015MNRAS.454.3771P} 
show that super-Eddington accretion may occur in highly-obscured environments, 
where radiation trapping is so efficient that photons are advected inward rather 
than being radiated away: thus, the effect of radiation pressure is dramatically reduced.} 
with radiative efficiency of 0.1, 
we can translate the accretion rate density $\dt{\rho}_\bullet$ 
into an {\it active} BH mass density $\rho_\bullet$:
\begin{equation}
\frac{\rho_\bullet} {M_\odot~{\rm Mpc^{-3}}} \approx4.1\times10^{7}  \frac{\dt{\rho}_\bullet} 
{M_\odot~{\rm yr^{-1}~Mpc^{-3} }}.
\end{equation}
Therefore, the critical active BH mass densities above which the XRB gives negative feedback 
for the DCBH formation are about $4\times10^3~M_\odot {\rm Mpc^{-3}}$, $4\times10^2~M_\odot {\rm Mpc^{-3}}$ and $4\times10^1~M_\odot {\rm Mpc^{-3}}$ for the three $N_{\rm H}$ models, respectively.

With an accretion rate density $\dt{\rho}_\bullet$ higher than the above value, the XRB effect becomes noticeable and $J_{\rm LW}^{\rm crit}$ increases with $\dt{\rho}_\bullet$. However, even with $\dt{\rho}_\bullet$ as high as the maximum allowed 
by the present-day XRB level, the DCBH can still form with feasible $J_{\rm LW}^{\rm crit}$.
For the BH1 model, the maximum allowed accretion rate density is $\approx0.22~M_\odot {\rm yr^{-1} Mpc^{-3}}$, 
corresponding to $\rho_\bullet =9.0\times10^6~M_\odot {\rm Mpc}^{-3}$. At this limit $J_{\rm LW}^{\rm crit}$ only increases to $\approx$ 80.
Under the same hypothesis, for the   BH2(3) model 
at the maximally allowed accretion rate density, 
$\dt{\rho}_\bullet = 0.034 (0.006)~M_\odot {\rm yr^{-1}Mpc^{-3}}$, 
we get $J_{\rm LW}^{\rm crit} \approx 170$ (390).

All these $J_{\rm LW}^{\rm crit}$ values are still much smaller than 
those for normal star-forming galaxies, i.e. $\sim 1000-10000$ 
\citep{2014MNRAS.445..544S,2015MNRAS.446.3163L,2016MNRAS.459.4209A}.
This indicates that, even with the XRB, 
the DCBHs are efficient radiation sources for triggering DCBHs in nearby halos.
Moreover, an accreting BH is usually much brighter than a galaxy, e.g. for a DCBH with mass $10^6~M_\odot$, the Eddington luminosity is $\approx1.3\times10^{44}~$erg s$^{-1}$, while a galaxy with our GAL spectrum and 10 $M_\odot$yr$^{-1}$ has bolometric luminosity $\approx3\times10^{42}$ erg s$^{-1}$. Hence, even if $J_{\rm LW}^{\rm crit}$ is the same, 
a BH radiation can exceeds $J_{\rm LW}^{\rm crit}$ more easily
and can suppress H$_2$ formation in a larger number of nearby metal-free halos \citep{2014MNRAS.440.1263Y}.
Furthermore, if a star-forming galaxy is the dissociation radiation source, it also produces strong 
soft X-rays, which would counteract the H$_2$ dissociation (see Sec. \ref{secXFG} below).
However it does not happen for the surrounded DCBHs since soft X-ray photons 
do not leak out from the envelope during the Compton-thick stage ($\sim100$ Myr).

Finally, we evaluate the appropriate distance from a halo to bear DCBH (``hatching halo'') 
to the radiation source (``source halo'').
If the hatching halo is close enough to the source, the radiation intensity exceeds 
$J_{\rm LW}^{\rm crit}$ and H$_2$ formation is suppressed.
However, if the distance among them is too small, the hatching halo could by 
tidally disrupted by gravity of the source halo. 
The tidal radius of the hatching halo is defined as \citep{1987gady.book.....B,2016ApJ...832..134C}  
\begin{equation}
r_{\rm TD} =\left(  \frac{M_{\rm hat}}{3M_{\rm sou}} \right)^{1/3}d,
\end{equation}
where $M_{\rm hat}$ and $M_{\rm sou}$ are the masses of the hatching halo 
and source halos, respectively, and $d$ the distance among them. 
The matter of the hatching halo outside the tidal radius 
would be removed by the tidal force. 

We conservatively require that for the hatching halo to keep its density without tidally disrupted, 
the tidal radius $r_{\rm TD}$ be larger than the virial radius $r_{\rm vir}$.
Assuming $M_{\rm hat} \approx M_{\rm sou}$, we obtain $d > \sqrt[3]{3}r_{\rm vir}=1.44 r_{\rm vir}$.
In Fig. \ref{fig_J21_vs_rho_acc}, we indicate the regions where the XRB from DCBHs 
is well below the present-day XRB constraints and the required $J_{\rm LW}^{\rm crit}$ can be provided by 
a source DCBH of $2\times10^5$ or $1\times10^6~M_\odot$ locating at the distance 
larger than  $d> 1.44 r_{\rm vir}$, where $r_{\rm vir}$ is evaluated for $T_{\rm vir}=10^4$K and 
$z=13$. 
Fig. \ref{fig_J21_vs_rho_acc} indicates that, for any DCBH spectrum considered, 
there is still a large parameter space for the DCBH formation without being tidally disrupted.

\subsubsection{X-rays from both the first galaxies and DCBHs}\label{secXFG}
Next, we also consider the XRB created by the first galaxies $J_X^{\rm gal}$ and add it 
to that by DCBHs $J_X^{\bullet}$. 

We assume the high-mass X-ray binaries (HMXBs) as the sources of XRB 
since they contribute most to the galactic X-ray radiation 
among three major contributors, i.e., HMXBs, 
low-mass X-ray binaries and hot interstellar medium 
(see e.g. \citealt{2014MNRAS.443..678P} for details). 
The X-ray SED of HMXBs follows the power law with index $\alpha_X$, 
with its normalization depending on the SFR: 
\begin{equation}
L_{X}^{\rm HMXB}(E)=\frac{(1-\alpha_X)L_{\rm band}({\rm SFR},z) }{E_2^{(1-\alpha_X)}-E_1^{(1-\alpha_X)}} E^{-\alpha_X},
\end{equation} 
where $L_{\rm band}$ is the X-ray luminosity in the energy range considered 
($E_1<E<E_2$) and we adopt the spectral index $\alpha_X=1.5$ in the following.
We use the fitting formula for the luminosity in 2 - 10 keV (below 2 keV the X-ray photons might be absorbed by host galaxy gas, therefore the above normalization would be not valid, see \citealt{2014MNRAS.443..678P}) by \citet{2013ApJ...762...45B}
\begin{align}
&{\log} \left( \frac{L_{\rm band}}{{\rm erg~s^{-1}}} \right)= \nonumber \\
&0.97{\log}(1+z)+0.65{\log}\left(0.67 \frac{{\rm SFR}}{{\rm M_\odot yr^{-1}}}\right)+39.80,
\label{eq_L_band}
\end{align}
and to model the scatter we assume a 0.5 dex variance for this relation.
Although the almost linear dependence on redshift, 
$L_{\rm band}\propto (1+z)^{0.97}$, is obtained from fitting to $z<4$ samples, 
we here extrapolate it to galaxies at higher redshifts. 
We have multiplied the SFR by a factor 0.67 \citep{2014ARA&A..52..415M} so that here the SFR
corresponds to Salpeter IMF with mass limits of 0.1 and 100 $M_\odot$, we adopted.
At $z=13$ and SFR$=1~M_\odot$yr$^{-1}$,
Eq. (\ref{eq_L_band}) gives log$(L_{\rm band})=40.8^{+0.5}_{-0.5}$.
Note that this brackets the fiducial value $40.5$ adopted in \citet{2006MNRAS.371..867F}.

The SFR is assumed proportional to halo mass growth rate:
\begin{equation}
{\rm SFR}(M,z)=f_*\frac{\Omega_b}{\Omega_m}\frac{M}{2\Delta t_{\rm SF}(M,z)},
\label{eq_halo_SFR}
\end{equation}
where $M$ is the halo mass, and 
the star formation time scale, $\Delta t_{\rm SF}$, is the time since the median formation time, 
which is defined as the time when the halo has collected half of its mass, 
to the considering redshift $z$ \citep{2007MNRAS.376..977G}, 
and $f_*$ is the star formation efficiency. 

The star formation efficiency $f_*$ is calibrated from the observed 
UV luminosity density $\rho_{\rm UV}$, which is 
derived from the UV luminosity functions of high-$z$ galaxies. 
By extrapolating the luminosity function in \citet{2015ApJ...803...34B} down to 
the absolute UV magnitude $-10$, \citet{2015ApJ...811..140B} has obtained 
$\rho_{\rm UV}$ as a function of redshift and it is 
$\rho_{\rm UV}=10^{26.2\pm0.2}$  erg s$^{-1}$Hz$^{-1}$ Mpc$^{-3}$ at $z=10$.
By equating this with the sum of contributions from all halos calculated theoretically:
\begin{equation}
\rho_{\rm UV}=\int_{M_{\rm min}}  l_{\rm UV}(\Delta t_{\rm SF}) 
{\rm SFR}(M,z) \frac{d {\mathcal N}}{dM}dM,
\label{eq_rhoUV}
\end{equation}
where $M_{\rm min}$ is the minimum halo mass that can form stars and for which 
we adopt the virial mass of $T_{\rm vir}=10^4$ K, 
$d{\mathcal N}/dM$ is the halo mass function \citep{1999MNRAS.308..119S,2001MNRAS.323....1S} 
and $l_{\rm UV}$ is the UV (at 1500 \AA) luminosity per unit SFR for continuous 
star formation mode with Salpeter IMF (0.1 - 100 $M_\odot$) and 
metallicity $Z=0.02~Z_\odot$, taken from {\tt STARBURST99}.
We then obtain $f_*=0.022^{+0.012}_{-0.008}$. 

The XRB from first galaxies at redshift $z$ can be calculated by
\begin{equation}
J_X^{\rm gal}(E,z)=(1+z)^3\int_z
e_X^{\rm gal}(E',z') {\rm e}^{-\tau_{\rm IGM}(E')} \frac{dr_{\rm p}}{dz'}dz',
\label{eq_JX_gal}
\end{equation}
where the X-ray emissivity 
\begin{align}
e_X^{\rm gal}(E^\prime,z)=  \frac{1}{4\pi} 
\int_{M_{\rm min}} f_{\rm esc, X}(E')  L_{\rm X}^{\rm HMXB}(E') 
\frac{d {\mathcal N}}{dM} dM.
\end{align}
The X-ray escape fraction $f_{\rm esc,X}$  depends on the neutral hydrogen, neutral helium and metal content of host galaxies,
\begin{align}
f_{\rm esc, X}(E)={\rm exp}\left[-(\sigma_{X,1}(E) N_{\rm HI}^{\rm gal}+\sigma_{X,2}(E) ZN_{\rm HI}^{\rm gal})\right],
\end{align}
where $N_{\rm HI}^{\rm gal}$ is the column number density of neutral hydrogen, 
$\sigma_{X,1}$ and $\sigma_{X,2}$ the synthesis photo-electric cross-section 
of neutral hydrogen and helium only, and of metal elements with  Solar abundance, respectively 
Both $\sigma_{X,1}$ and $\sigma_{X,2}$ have been converted to those in units 
per hydrogen atom \citep{1992ApJ...400..699B}.

The neutral hydrogen column density in first galaxies is largely unknown. 
\citet{2016ApJ...825....7L} provide $z$-dependent expressions for the X-ray luminosity 
in both 0.5 - 2 keV and 2 - 10 keV bands. 
By assuming that the 2 - 10 keV luminosity approximates the intrinsic one, while that 
of 0.5 - 2 keV is attenuated one, and comparing these two, we derive 
$N_{\rm HI}^{\rm gal}\sim3\times10^{21}$ - $3\times10^{22}$ cm$^{-2}$ 
for SFR $\sim10^{-3}$ - $50~M_\odot$yr$^{-1}$ at $z=13$. 
Although such value should be only considered as an educated guess, 
we nevertheless assume a fiducial value log$N_{\rm HI}^{\rm gal}=22$ with $\pm0.5$ dex scatter, 
and use it for all $z\ge13$. 

\begin{figure}
\centering{
\includegraphics[scale=0.4]{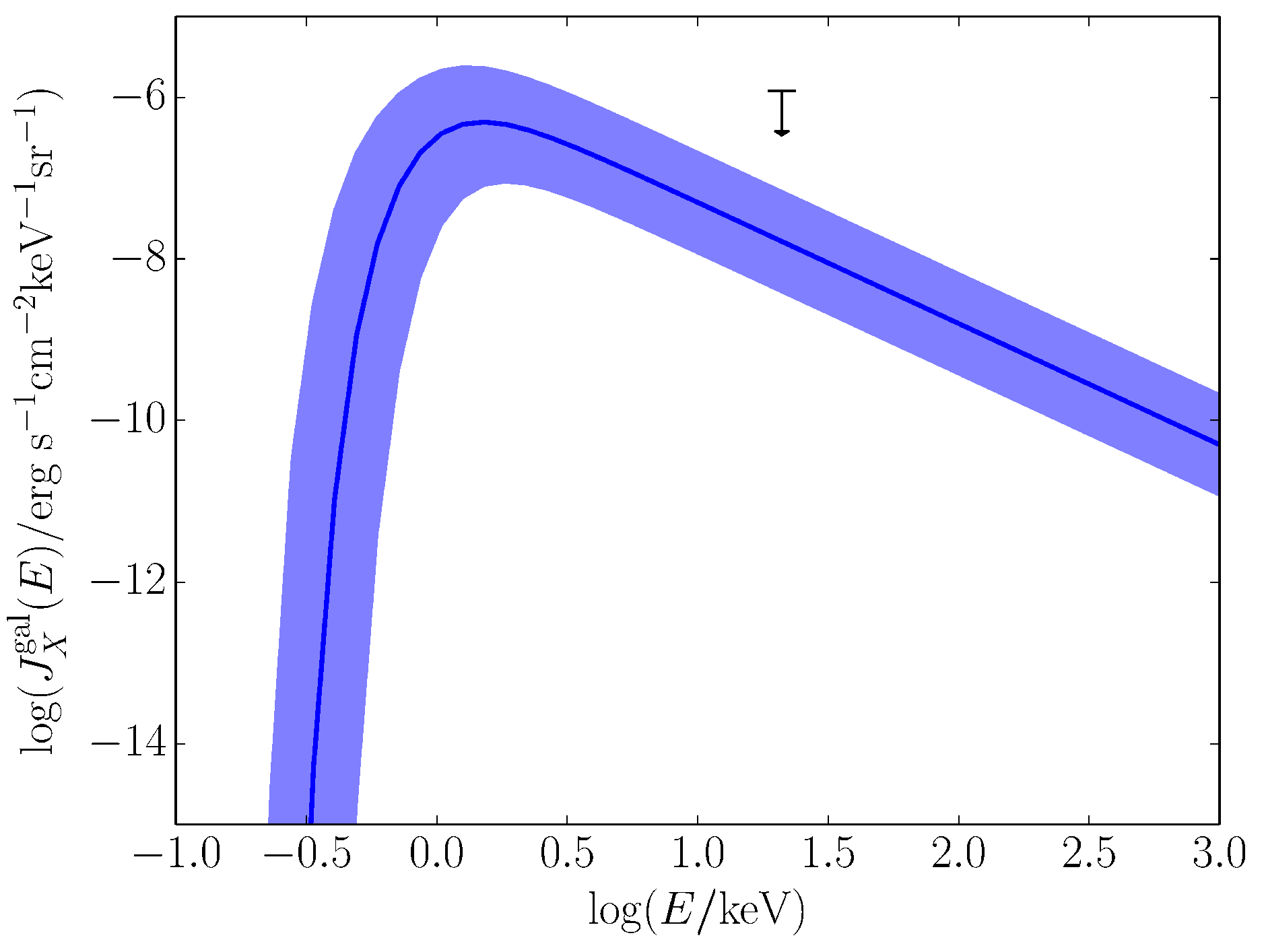}
\caption{The $z=13$ XRB from first galaxies. Fiducial model with $f_*=0.022$, log$N_{\rm HI}^{\rm gal}=22$ and ${\rm log}L_{\rm band}=0.93{\rm log}(1+z)+0.65{\log}(0.67\times{\rm SFR})+39.80$ is plotted by solid line. The variance  corresponding to $f_*=0.022^{+0.012}_{-0.008}$, log$N_{\rm HI}=22^{+0.5}_{-0.5}$ and ${\rm log}L_{\rm band}=0.93{\rm log}(1+z)+0.65{\log}(0.67\times{\rm SFR})+39.80\pm0.5$  is shown with a shaded region.
} 
\label{fig_J_X_first_galaxies}
}
\end{figure}

\begin{figure*}
\centering{
\includegraphics[scale=0.8]{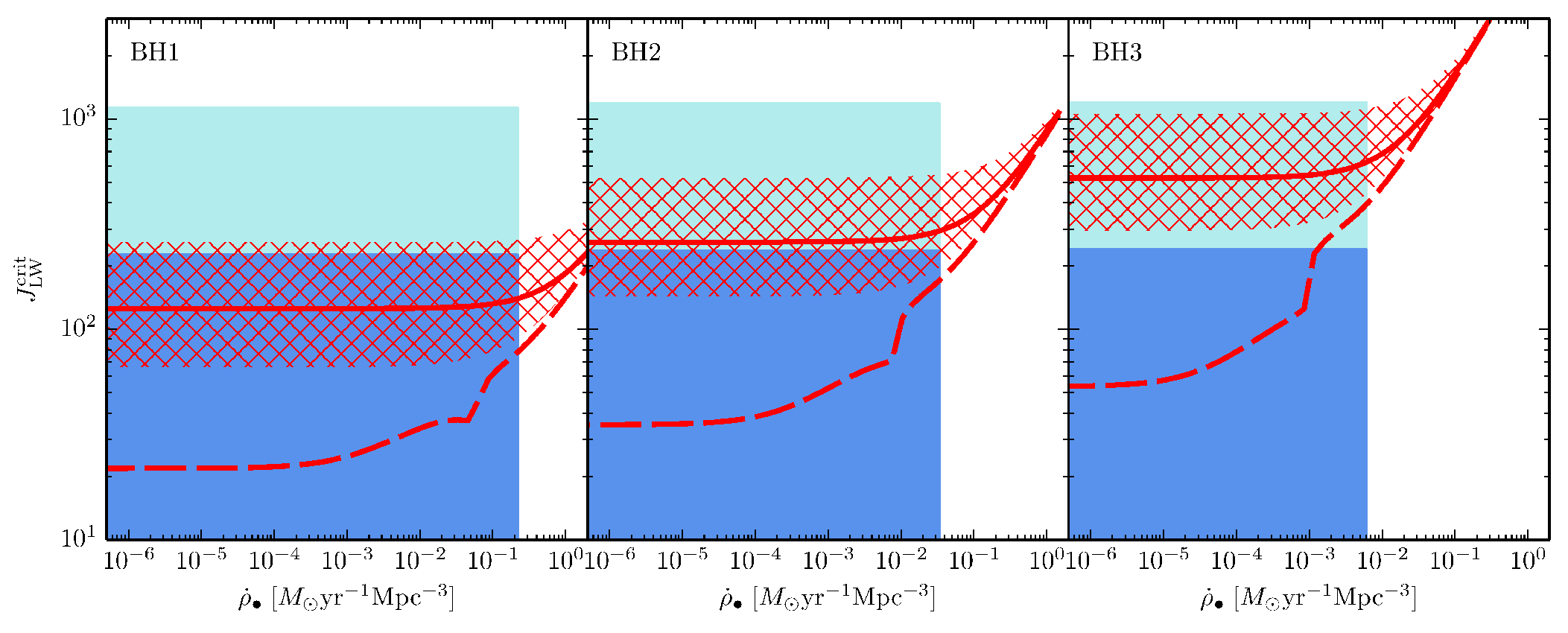}
\caption{The $J_{\rm LW}^{\rm crit}$ against $\dot{\rho}_\bullet$, XRB from first galaxies is included as well. The shaded regions are allowed space by present-day XRB constraints and gravitationally disruption, see Fig. \ref{fig_J21_vs_rho_acc}. Solid curves correspond to fiducial parameters of first galaxies in Fig. \ref{fig_J_X_first_galaxies} while hatched regions are variance. For reference we also plot the $J_{\rm LW}^{\rm crit}$ curves in the absence of XRB from first galaxies by dashed lines.
} 
\label{fig_J21_vs_rho_acc_with_first_galaxies}
}
\end{figure*}

In Fig. \ref{fig_J_X_first_galaxies}, we show the XRB from first galaxies at $z=13$ with the variance. 
They contribute $\sim0.3\%-6\%$ to the unresolved present-day XRB at 1.5 keV. 
The total XRB is the sum of $J_X^{\rm gal}$ and the DCBH contribution $J_X^{\bullet}$.
Recall that the latter is a function of $\dt\rho_{\bullet}$.
For this XRB, we re-calculate $J_{\rm LW}^{\rm crit}$ for each $\dt\rho_{\bullet}$. 
The results are plotted in Fig. \ref{fig_J21_vs_rho_acc_with_first_galaxies}. 
Comparing with the values with DCBH XRB only, the $J_{\rm LW}^{\rm crit}$ is boosted by about factors of 2 - 20.
This indicates that the XRB from first galaxies 
may have significant negative influence on the DCBH formation 
although DCBH formation is still feasible. 

\subsection{The number density of DCBHs}
 
The formation rate of DCBHs is
\begin{equation}
\frac{dn_{\rm DCBH}}{dz}\sim \frac{dn_{\rm cool}}{dz}P_{\rm DCBH},
\label{eq_dn_DCBH}
\end{equation}
where $dn_{\rm cool}/dz$ is the formation rate of atomic-cooling halos, $P_{\rm DCBH}$ is the probability that a newly-formed atomic-cooling halo matches the DCBH formation criteria. Here we simply assume it to be equal to the probability that this halo is located in a radiation field above the critical threshold. We ignore the metal enrichment and the radiative feedback \citep{2014MNRAS.440.1263Y,2014MNRAS.445..686J}. This probability depends on the spatial distribution of the already-formed nearby galaxies and DCBHs. Atomic-cooling halos are assumed to host either star-forming galaxies with SFR as in Eq. (\ref{eq_halo_SFR}) (here we adopt $f_*=0.1$), or DCBHs. Using the two-point correlation function which describes the spatial distribution of their host halos, we calculate $P_{\rm DCBH}$ by the Monte Carlo simulations as in \citet{2008MNRAS.391.1961D,2014MNRAS.442.2036D,2014MNRAS.440.1263Y}. 

We solve the Eq. (\ref{eq_dn_DCBH}) from $z = 25$, assuming that the initial number density of DCBHs at this redshift is zero.  For simplicity we assume all DCBHs to have the same luminosity as a $5\times10^5~M_\odot$ black hole.

\begin{figure}
\centering{
\includegraphics[scale=0.4]{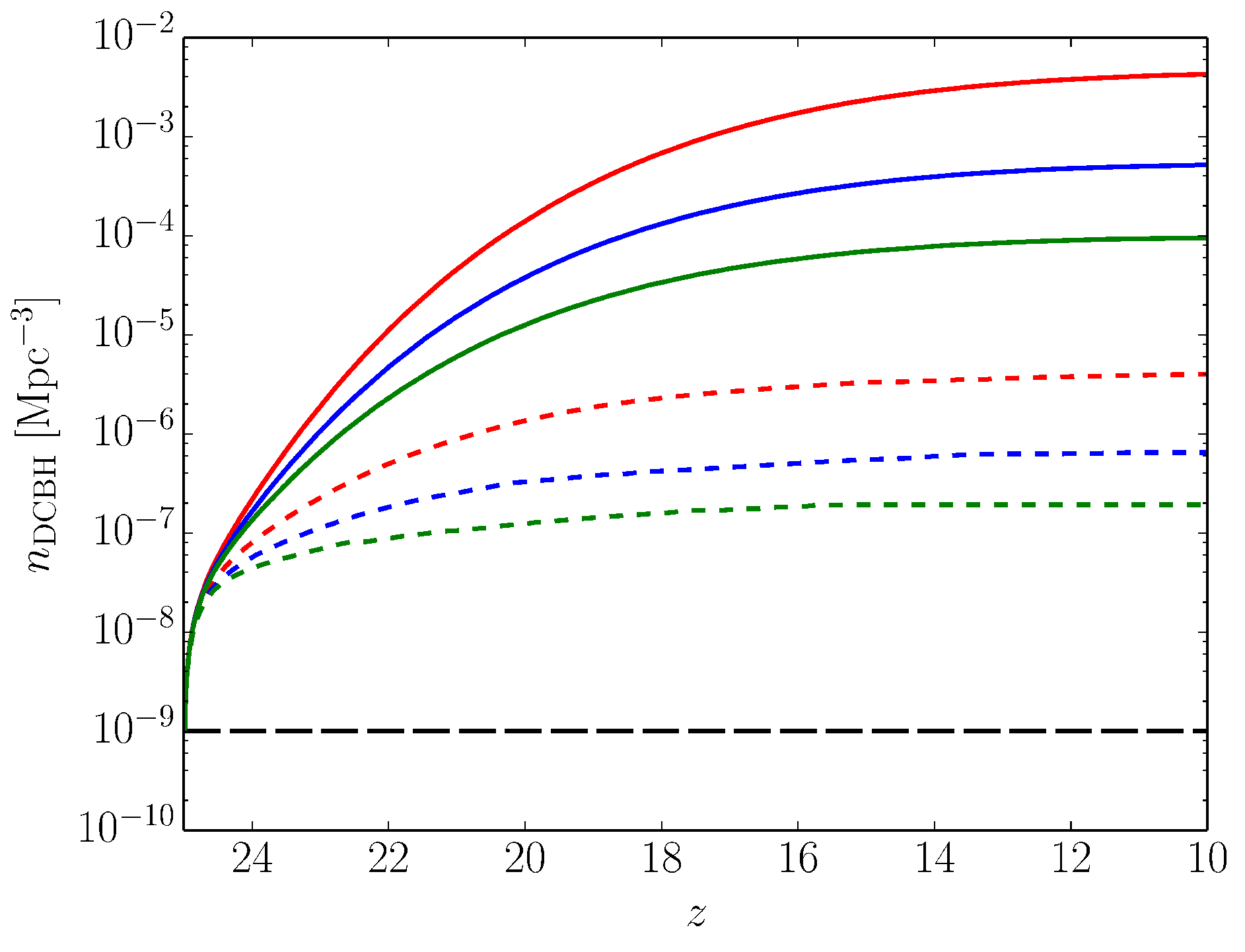}
\caption{The evolution of the number density of DCBHs when different SEDs and critical field strengths are adopted. Solid (dashed) lines are for the critical field strengths without (with) XRB from first galaxies.  From top to bottom the three solid lines correspond to BH1 SED with $J_{\rm LW}^{\rm crit} = 22$, BH2 SED with $J_{\rm LW}^{\rm crit}=35$ and BH3 SED with $J_{\rm LW}^{\rm crit}=54$ respectively.   The three dashed lines correspond to BH1 SED with $J_{\rm LW}^{\rm crit}=130$, BH2 SED with $J_{\rm LW}^{\rm crit}=260$ and BH3 SED with $J_{\rm LW}^{\rm crit}=520$ respectively, see Fig. \ref{fig_J21_vs_rho_acc_with_first_galaxies}. The horizontal dashed line refers to the approximate number density of observed SMBHs at $z\gsim6$,  $n_{\rm SMBH}(z\gsim6)\sim10^{-9}$~Mpc$^{-3}$.} 
\label{fig_n_DCBH}
}
\end{figure}

In Fig. \ref{fig_n_DCBH} we show the evolution of the DCBH number density for different SEDs and critical field strengths. In all models the final number density of DCBHs is fairly above the observed number density of SMBHs at $z\gsim6$, $n_{\rm SMBH}(z\gsim6)\sim10^{-9}$ Mpc$^{-3}$. In some models $n_{\rm DCBH}$ is even above the $\sim\frac{M_{\rm SMBH}}{M_{\rm DCBH}}n_{\rm SMBH}\sim \frac{10^9}{10^5}10^{-9}=10^{-5}$ Mpc$^{-3}$. Therefore DCBHs can perform as the original seeds of SMBHs. However, we clarify that there are plenty of factors that can influence the $n_{\rm DCBH}$, for example the number of DCBHs initially formed by radiation from galaxies; the minimum separation between the collapsing cloud and the triggering source (we adopt 1.44 $r_{\rm vir}$ as discussed in Sec. \ref{XRB_DCBH}\footnote{If the triggering source is a galaxy, the minimum separation must be also larger than the ionized bubble around  the galaxy. The ionized bubble radius is estimated by the equilibrium between the ionization rate and the recombination rate,
\begin{equation}
\frac{4\pi}{3}r_{\rm B}^3n_{\rm H}^2\alpha_{\rm B}=f_{\rm esc}\dot{Q}_{\rm H},
\end{equation}
where $n_{\rm H}$ is the mean gas density in the ionized bubble, $\alpha_{\rm B}$ is the recombination coefficient, $f_{\rm esc}$ and $\dot{Q}_{\rm H}$ are the escape fraction and production rate of the ionizing photons respectively. From \citet{2014MNRAS.444L.105P}, the typical gas density near the galaxy is $\sim 10 - 10^{2.5}$ times the mean density of the Universe. However, minihalos and dense filaments may boost this factor. Because of the uncertainties on $n_{\rm H}$ and $f_{\rm esc}$, we do not include $r_{\rm B}$ in our calculation in this subsection.
}); 
 the metal enrichment and radiative feedback (ignored here); the mass distribution of DCBHs (we assume a fixed mass $5\times10^{5}~M_\odot$). Therefore what we present here is not considered as a prediction, instead it is only a proof-of-concept calculation to show how the DCBH number density should be estimated. A conclusive prediction must be based on models that treat the above factors more realistically.

\section{Conclusions and discussion}\label{conclusions}

We have investigated the critical specific intensity of the DCBH radiation that could suppress $\rm H_2$ formation and cooling in a nearby collapsing gas cloud by one-zone calculations. If the radiation from the emitting DCBH is higher than the critical intensity, we expect that a new DCBH forms in the gas cloud. We used three realistic DCBH SED models from numerical simulations with initial column number density for the absorbing envelope $N_{\rm H} = 1.3\times10^{25}$ cm$^{-2}$, $N_{\rm H} = 8.0\times10^{24}$ cm$^{-2}$ and $N_{\rm H} = 5.0\times10^{24}$ cm$^{-2}$, respectively.   

We have found that:
\begin{itemize}
\item DCBH spectra are very effective at photo-detaching H$^-$, a catalyst species for H$_2$ formation. Depending on the obscuring gas column density,  the ratio between H$^-$ photo-detachment and H$_2$ photo-dissociation rates for the DCBH with  $N_{\rm H} = 1.3\times10^{25}$ cm$^{-2}$ is $k_{\rm H^-, det}^{\rm BH1}/k_{\rm H_2,dis}^{\rm BH1}=4.1\times10^4$, i.e. comparable to that of a blackbody spectrum with effective temperature $10^4$~K which is $k_{\rm H^-, det}^{\rm BB}/k_{\rm H_2,dis}^{\rm BB}=7.2\times10^4$. For the other two DCBH SEDs the $k_{\rm H^-, det}^{\rm BH2}/k_{\rm H_2,dis}^{\rm BH2}=1.4\times10^4$ and $k_{\rm H^-, det}^{\rm BH3}/k_{\rm H_2,dis}^{\rm BH3}=4.8\times10^3$, still much higher than a typical star-forming galaxy which only yields $k_{\rm H^-, det}^{\rm GAL}/k_{\rm H_2,dis}^{\rm GAL}=44$.

\item Ignoring the effect of X-rays, the critical field intensity to suppress H$_2$ formation for the three DCBH SEDs is $J_{\rm LW}^{\rm crit}\approx22$, 35 and  54, respectively. Note that this is similar to a blackbody radiation field with effective temperature $10^4$ K, ($J_{21}^{\rm crit} \approx 30$), but much lower than for a typical star-forming galaxy spectrum ($J_{\rm LW}^{\rm crit}\approx700$). Hence, an emitting DCBH can trigger the formation of DCBH in a nearby collapsing gas cloud much more efficiently than galaxies. 
 
\item If an XRB produced by previously formed DCBH is present, it may promote H$_2$ formation and increase $J_{\rm LW}^{\rm crit}$. For the  $N_{\rm H} = 1.3\times10^{25}$ cm$^{-2}$ model, if the accretion rate density is $\dt{\rho}_\bullet\lsim 10^{-4}~M_\odot$yr$^{-1}$Mpc$^{-3}$, XRB plays a negligible role. However, even if $\dt{\rho}_\bullet$ reaches the maximum value allowed by the present-day XRB level, i.e. $\sim0.22~M_\odot$ yr$^{-1}$Mpc$^{-3}$, $J_{\rm LW}^{\rm crit}$ only increases to $\approx 80$.
For the $N_{\rm H} = 8.0(5.0)\times10^{24}$ cm$^{-2}$ SED models, XRB effect is negligible when $\dt{\rho}_\bullet\lsim 10^{-5}(10^{-6})~M_\odot$yr$^{-1}$Mpc$^{-3}$, and the $J_{\rm LW}^{\rm crit}$ increases to $\approx170$(390) at the maximum $\dt{\rho}_\bullet$ allowed by the present-day XRB level, which is $\approx0.034(0.006)~M_\odot$ yr$^{-1}$Mpc$^{-3}$. However, if the additional but uncertain XRB contribution from first galaxies is included,  it may modify the result by factors in the range $\sim$2 - 20.
\end{itemize}

Finally, it is worth noting that several works, for instance \citet{ 2010MNRAS.402.1249S}, \citet{2014MNRAS.443.1979L} and \citet{2015MNRAS.446.3163L}, have shown that the $J_{\rm LW}^{\rm crit}$ derived from 3D simulations is almost $\sim10-100$ times higher than from one-zone calculations,  ``{\it due to the inability of one-zone models to simulate shocks, collapse dynamics and hydrodynamical effects}" \citep{2015MNRAS.446.3163L}. Moreover, they also found that the $J_{\rm LW}^{\rm crit}$ varies from halo to halo. Under these circumstances, we might be significantly underestimating $J_{\rm LW}^{\rm crit}$.  Further numerical work will be necessary to clarify this point.

\acknowledgments

B.Y. is supported by the CAS Pioneer Hundred Talents (Young Talents) Program. F.P. acknowledges the NASA-ADAP grant MA160009. This work is supported in part by the Grant-in-Aid from the Ministry of Education, Culture, Sports, Science and Technology (MEXT) of Japan (25287040 KO).

\newpage

\appendix 

\section{The optical depths for H$_2$, H$^{-}$, and H$_2^+$}\label{optical_depth}
Here we check the validity of the optically thin assumption made for the H$^-$ detachment radiation and  for the H$_2^+$ dissociation radiation. For H$^-$, the mean optical depth of the gas cloud is
\begin{equation}
\mean{\tau_{\rm H^-,det}}=\lambda_J/2 \mean{\sigma_{\rm H^-,det}}n_{\rm H^-},
\end{equation}
where the mean cross-section is 
\begin{equation}
\mean{\sigma_{\rm H^-,det}} =k_{\rm H^-, det}\left(\int_{\nu_{0.755}}^{\nu_{13.6}} \frac{4\pi J(\nu)}{h \nu} d\nu\right)^{-1}.
\end{equation}
This mean cross-section  only depends on the spectral shape of the external radiation field, however is independent of its strength which appears in both numerator and denominator. For the BB spectrum, $\mean{\sigma_{\rm H^-,det}}=2.8\times10^{-17}$ cm$^2$. Similarly, we can also calculate the mean cross-section for H$_2^+$, for which we get $ \mean{\sigma_{\rm H_2^+,dis}} =7.0\times10^{-19}$ cm$^{-2}$, and the corresponding mean optical depth $\mean{\tau_{\rm H_2^+,dis}}$.
 
\begin{figure}
\centering{
\includegraphics[scale=0.4]{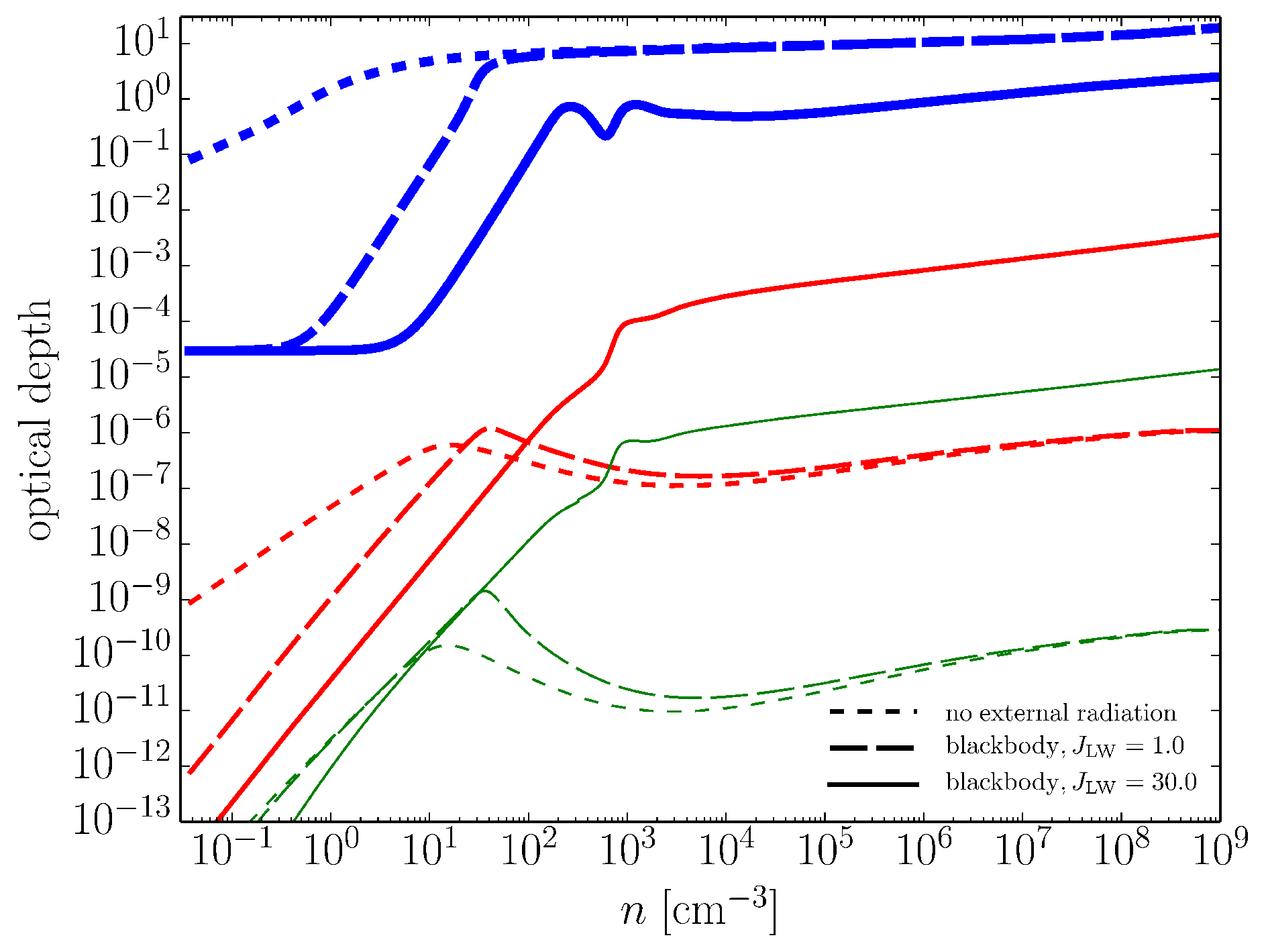}
\caption {The optical depth against $n$, thick lines are $-{\rm log}(f_{\rm sh})$ for H$_2$ self-shielding, medium and thin lines are for the H$^-$ and H$_2^+$ respectively.}
\label{fig_tau_BB1e4K}
}
\end{figure}

In Fig. \ref{fig_tau_BB1e4K} we plot $\mean{\tau_{\rm H^-,det}}$, $\mean{\tau_{\rm H_2^+,dis}}$ and $-{\rm log}(f_{\rm sh})$ (considered as the mean optical depth to LW radiation) respectively against the $n$ for different external radiation field strengths. This figure shows that, since the optical depth to H? detachment radiation and to H+2 dissociation radiation are always much smaller, it is then safe to ignore them.  On the other hand, even when $J_{\rm LW}\gsim J_{\rm LW}^{\rm crit}$, at $n\gsim 10^3$ cm$^{-3}$ the self-shielding effect to LW radiation is not negligible.

\section{Parameter windows for enhanced DCBH formation probability}\label{window}

In our fiducial model we adopt the following initial setup: $z_{\rm ta} = 30.6$; adiabatic IGM temperature; ionization fraction $10^{-4}$, H$_2$ fraction $10^{-6}$. In this section we investigate the influence of alternative initial setups on DCBH formation.

We first check that for BH1 radiation field, using $z_{\rm ta} = 16$  while keeping other parameters fixed to the fiducial ones, we get $J_{\rm LW}^{\rm crit}\approx20$. As for the fiducial setup for which we get $J_{\rm LW}^{\rm crit}\approx22$, we conclude that DCBH formation is insensitive to cosmic epoch at which the process takes place.

We then take into account pre-heating of gas by X-rays and shocks before it collapses. In such case at $z=z_{\rm ta}$ the gas temperature is higher than the adiabatic IGM temperature. We set the initial gas temperature $T_0 = 30$, 50, 100  and 1000 K respectively, keeping other initial setups the same as our fiducial model, and find that $J_{\rm LW}^{\rm crit}$ increases with increasing $T_0$: $J_{\rm LW}^{\rm crit} \approx39$, 49, 50 and 51 for $T_0 = 30$, 50, 100 and 1000 K respectively. When $T_0 \gsim 100$ K the $J_{\rm LW}^{\rm crit}$ almost does no longer increase. We conclude that ignoring the gas pre-heating may underestimate the critical field intensity by a factor at most $\approx2$.

\begin{figure}
\centering{
\includegraphics[scale=0.4]{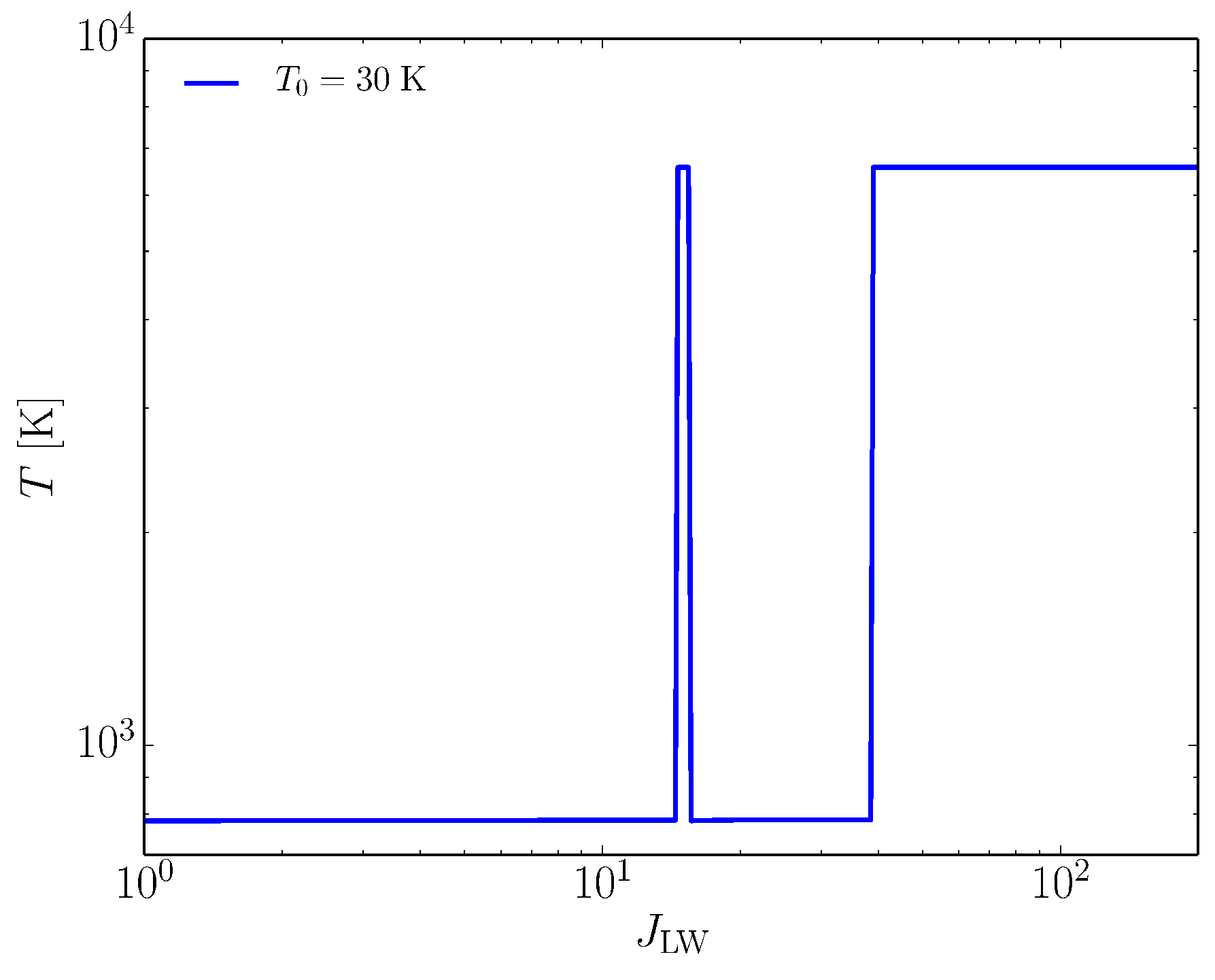}
\caption {For BH1 radiation field, the gas temperature at $n\sim 10^9$ cm$^{-3}$ vs. $J_{\rm LW}$  when gas has initial temperature $T_0 = 30$ K.}
\label{fig_Tfinal_vs_J21}
}
\end{figure}

Interestingly, we find that in the model with $T_0 = 30$ K, H$_2$ formation and cooling could also be suppressed even when the radiation field strength is smaller than $J_{\rm LW}^{\rm crit}\approx39$, as long as it is in a narrow range $14.5\lsim J_{\rm LW}\lsim15.5$, as shown by Fig. \ref{fig_Tfinal_vs_J21} where we plot the gas temperature at the final step of our calculations ($n\sim10^9$ cm$^{-3}$) against $J_{\rm LW}$. To investigate the physics behind this interesting phenomenon, we look through the evolution track and find that when $J_{\rm LW}$ is in this range, the gas almost simultaneously reaches the maximum temperature $\sim10^4$ K and the critical density for H-H$_2$ collisional dissociation, $n_{\rm cr,H}$ ( see \citealt{1996ApJ...461..265M}). When gas number density is higher than $n_{\rm cr,H}$, H$_2$ is dissociated via the ${\rm H_2 + H \xrightarrow{k_{13}} 3H}$ reaction with reaction rate $k_{13}$ approximately proportional to ${\rm exp}(-5\times10^4/T)$. It has strong temperature dependence hence the H$_2$ is dissociated very efficiently at the maximum gas temperature, resulting in suppressed H$_2$ cooling. If, however, the $J_{\rm LW}$ is smaller than $\sim$14.5, the gas cools earlier and never has a chance to reach the temperature $\sim10^4$ K, and the collision dissociation is not efficient because of the low temperature; or if the $J_{\rm LW}$ is larger than $\sim$15.5, the radiation reduces (but not fully suppresses) H$_2$ cooling and the gas temperature reaches $10^4$ K before the gas density reaches $n_{\rm cr,H}$, therefore the H$_2$ cannot be dissociated efficiently via H-H$_2$ collisions because of the low density, and then the gas temperature decreases soon. The same phenomenon also exists in the $T_0 = 50$ K model, except that the feasible  field intensity range is even narrower. However, it does not exist in our fiducial model where $T_0 = 19.8$ K, and the model with $T_0 > 50$ K.

Although the physics mechanism is interesting, the above phenomenon may hardly occur in more realistic cases in which the strength of the external radiation varies with time. However, this mechanism could work in other situations and decrease the critical field intensity, see below. 

\begin{figure}
\centering{
\includegraphics[scale=0.4]{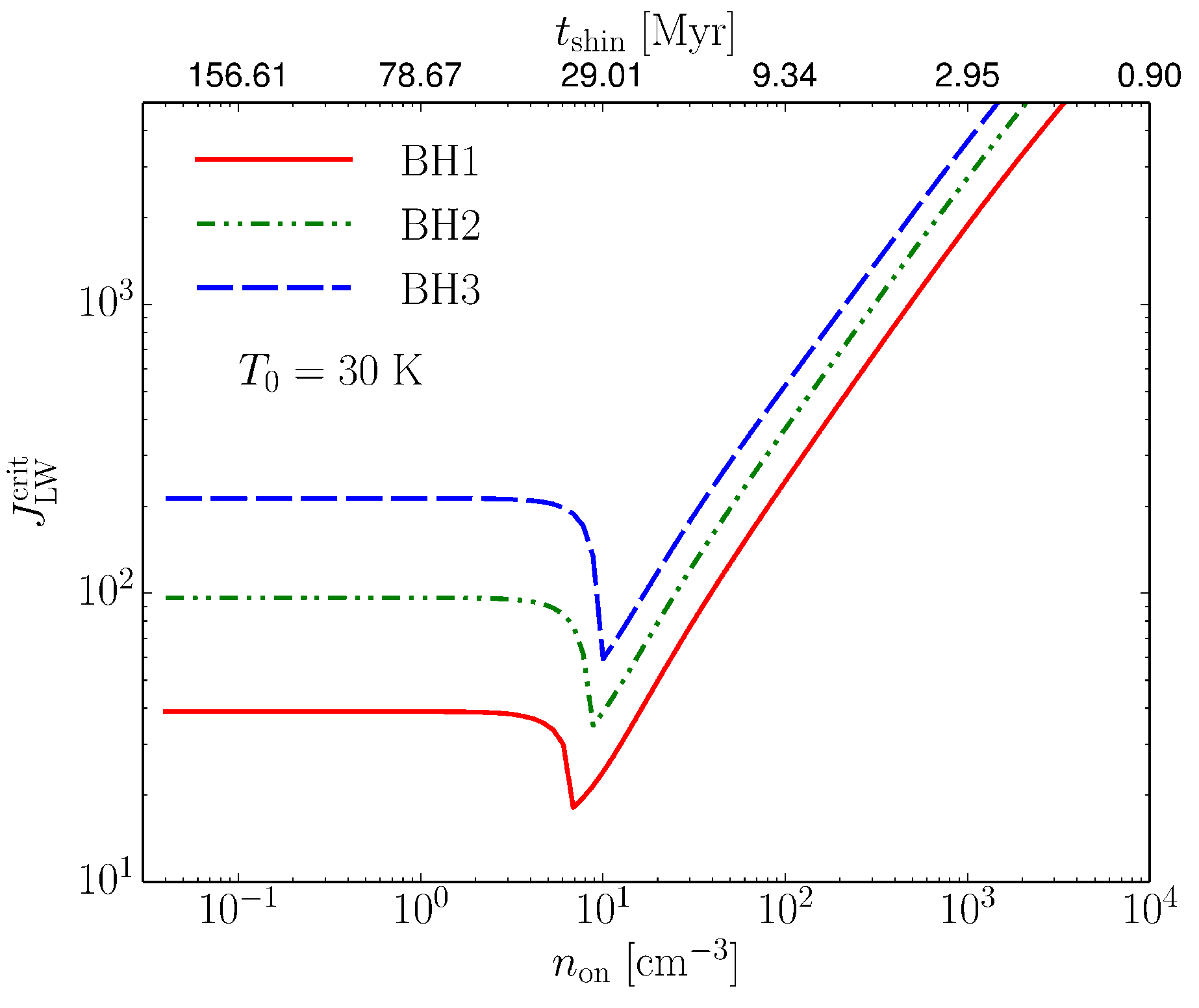}
\caption{Same to Fig. \ref{fig_J21_ntot_turnon} however $T_0 = 30$ K. }
\label{fig_J21_ntot_turnon_T0}
}
\end{figure}

We re-calculate the $J_{\rm LW}^{\rm crit}$ - $n_{\rm on}$ relation for $T_0 = 30$ K, the results are plotted in Fig. \ref{fig_J21_ntot_turnon_T0}. We find the $J_{\rm LW}^{\rm crit}$ does not increase monotonically with $n_{\rm on}$. Instead, for BH1 (BH2, BH3) radiation field when $2~{\rm cm}^{-3} \lsim n_{\rm on} \lsim 15~{\rm cm}^{-3}$ ($2~{\rm cm}^{-3} \lsim n_{\rm on} \lsim 25~{\rm cm}^{-3}$, $2~{\rm cm}^{-3} \lsim n_{\rm on} \lsim 35~{\rm cm}^{-3}$),  $J_{\rm LW}^{\rm crit}$ is even smaller than models where the external radiation field irradiates the collapsing gas cloud  longer. For example, if the external radiation is switched on at $\sim 7$ cm$^{-3}$,  $\rm H_2$ suppression is achieved with an intensity as low as $J_{\rm LW}^{\rm crit}\sim 18$ for BH1 radiation field.
We clarify here that the $J_{\rm LW}^{\rm crit}$ is not the left boundary of the narrow peak in Fig. \ref{fig_Tfinal_vs_J21}, instead it is the left boundary of the plateau, i.e. for all $J_{\rm LW} \ge J_{\rm LW}^{\rm crit}$ the H$_2$ formation and cooling will be suppressed.

\begin{figure}
\centering{
\includegraphics[scale=0.4]{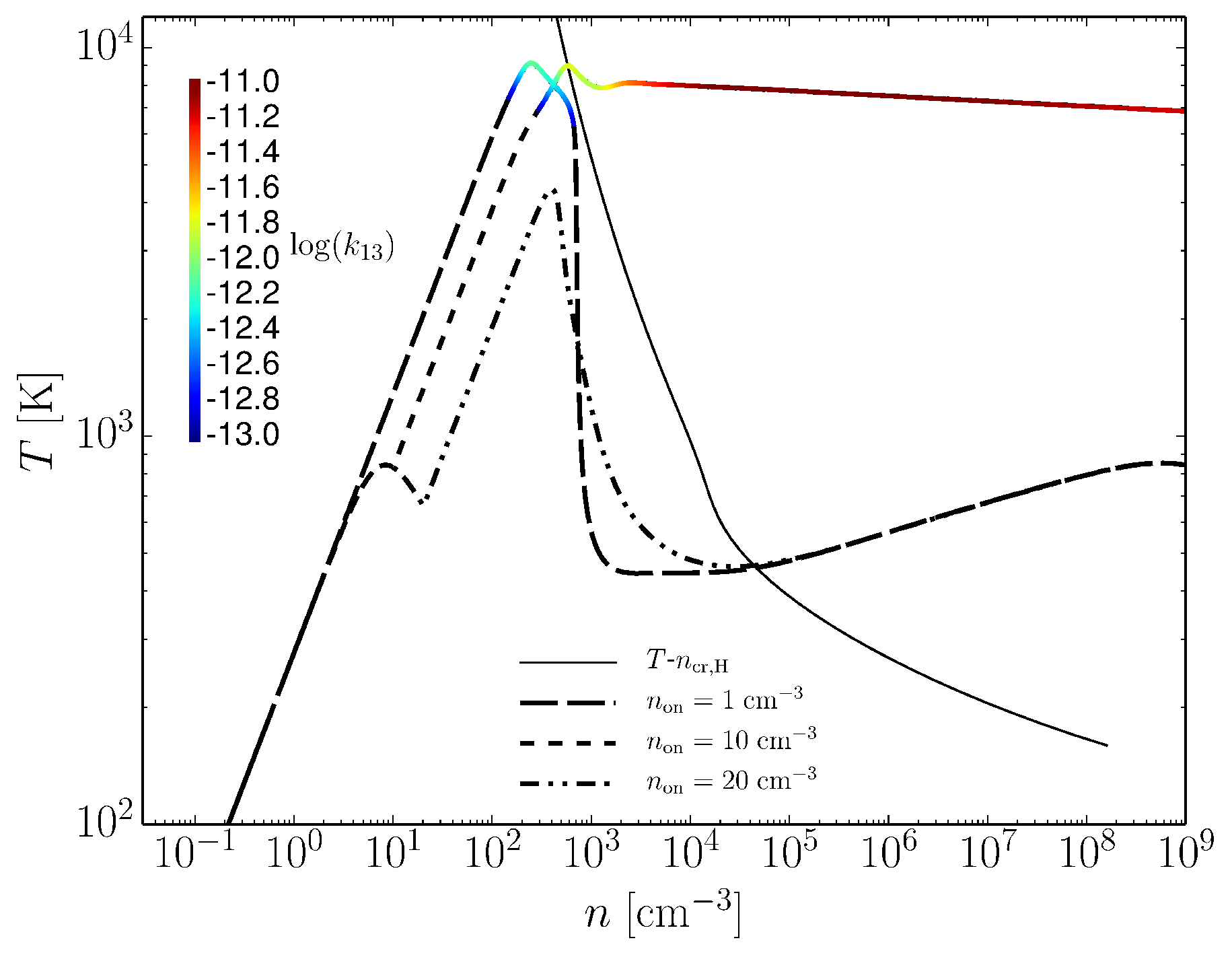}
\caption{The temperature track if the external BH1 radiation field switched on at  total number density 1, 10 and 20 cm$^{-3}$ respectively.  In the temperature we color the segment where the H$_2$ collisional dissociation rate $k_{13} > 10^{-13}$ cm$^3$ s$^{-1}$. In the same panel we also plot the $T$-$n_{\rm cr,H}$ curve.}
\label{fig_Tgas_and_ncrH}
}
\end{figure}

The physics of this  phenomenon is same to the above. Given $J_{\rm LW}=30$, in Fig. \ref{fig_Tgas_and_ncrH} we plot the the temperature tracks for $n_{\rm on} = (1, 10, 20)$ cm$^{-3}$. We also plot the $T$-$n_{\rm cr,H}$ curve in the same panel.  Note that $J_{\rm LW} = 30$ is smaller than the $J_{\rm LW}^{\rm crit}$ for $n_{\rm on} = 1$ and 20 cm$^{-3}$, but larger than the $J_{\rm LW}^{\rm crit}$ for $n_{\rm on} = 10$ cm$^{-3}$.  As seen from Fig. \ref{fig_Tgas_and_ncrH}, for $n_{\rm on} = 1$ cm$^{-3}$ and 20 cm$^{-3}$ cases, the temperature track and the $T$-$n_{\rm cr,H}$ curve intersect at $T\sim500$ K.  However for the $n_{\rm on} = 10$ cm$^{-3}$ case, the intersection is near 10000 K. Recall that at $\gsim n_{\rm cr,H}$ the reaction rate $k_{13} $ approximately proportional to ${\rm exp}(-5\times10^4/T)$. Hence, only if the temperature track and the $T$-$n_{\rm cr,H}$ curve intersect at high temperatures the $\rm H_2$ can be rapidly dissociated by H-H$_2$ collisions, see the colored segment of each curve where the $k_{13}$ is shown. In such a situation the H$_2$ formation and cooling suppression by a weaker external radiation is possible. Moreover, when $n_{\rm on} \gsim 100$ cm$^{-3}$, the $J_{\rm LW}^{\rm crit}$ increases rapidly because otherwise the previously formed H$_2$ has already cooled the gas to a low temperatures. We note that from $n\sim2$ cm$^{-3}$ to $\sim15$ cm$^{-3}$ only about $\sim 35$ Myr time elapses. It is still a short time-scale opportunity for DCBH to form in a weaker external radiation field.

\section{Reaction rates and cooling functions}\label{reactions}

Tables \ref{tb_reactions}$-$\ref{tb_coolings} list the rates of reactions, and heating and cooling functions adopted in this work.
 
\LongTables

\begin{deluxetable}{lllll} 
\tablecaption{rates of reactions adopted in this work.}
\tablehead{\colhead{Ordinal} &\colhead{Reaction} & \colhead{Rate\tablenotemark{a,b}  } &  & \colhead{units}} 
 \startdata

(1)& ${\rm H}+{\rm e}^- \rightarrow      {\rm H}^+ + 2{\rm e}^-$    &$ k_1={\rm exp}[-32.71396786375 $ && cm$^3$ s$^{-1}$ \\
&&~~~~~~~~~~~~~$+13.53655609057({\rm ln}T_{\rm eV}) $ \\
&&~~~~~~~~~~~~~$-5.739328757388({\rm ln}T_{\rm eV})^2 $\\
&&~~~~~~~~~~~~~$+1.563154982022({\rm ln}T_{\rm eV})^3$ \\
&&~~~~~~~~~~~~~$-0.2877056004391({\rm ln}T_{\rm eV})^4$\\
&&~~~~~~~~~~~~~$+0.03482559773736999({\rm ln}T_{\rm eV})^5$ \\
&&~~~~~~~~~~~~~$-0.00263197617559({\rm ln}T_{\rm eV})^6$\\
&&~~~~~~~~~~~~~$+0.0001119543953861({\rm ln}T_{\rm eV})^7$ \\
&&~~~~~~~~~~~~~$-2.039149852002\times10^{-6}({\rm ln}T_{\rm eV})^8]$ \\ \\
       
(2)&$ {\rm He}+{\rm e}^- \rightarrow {\rm He}^++ 2{\rm e}^-$ & $k_2={\rm exp}[-44.09864886561001 $ && cm$^3$ s$^{-1}$  \\
&&~~~~~~~~~~~~~$   +23.91596563469({\rm ln}T_{\rm eV}) $ \\
&&~~~~~~~~~~~~~$ -10.75323019821({\rm ln}T_{\rm eV})^2 $\\
&&~~~~~~~~~~~~~$+3.058038757198({\rm ln}T_{\rm eV})^3$ \\
&&~~~~~~~~~~~~~$-0.5685118909884001({\rm ln}T_{\rm eV})^4 $\\
&&~~~~~~~~~~~~~$+0.06795391233790001({\rm ln}T_{\rm eV})^5$ \\
&&~~~~~~~~~~~~~$-0.005009056101857001({\rm ln}T_{\rm eV})^6 $\\
&&~~~~~~~~~~~~~$ +0.0002067236157507({\rm ln}T_{\rm eV})^7$ \\
&&~~~~~~~~~~~~~$-3.649161410833\times10^{-6}({\rm ln}T_{\rm eV})^8]$  \\ \\

(3)& ${\rm H}^+ + {\rm e}^- \rightarrow {\rm H}+\gamma $ &  $k_3={\rm exp}[-28.61303380689232 $ && cm$^3$ s$^{-1}$\\
&&~~~~~~~~~~~~~$-0.7241125657826851({\rm ln}T_{\rm eV})$ \\
&&~~~~~~~~~~~~~$-0.02026044731984691({\rm ln}T_{\rm eV})^2 $\\
&&~~~~~~~~~~~~~$-0.002380861877349834({\rm ln}T_{\rm eV})^3$\\
&&~~~~~~~~~~~~~$-0.0003212605213188796({\rm ln}T_{\rm eV})^4$\\
&&~~~~~~~~~~~~~$-0.00001421502914054107({\rm ln}T_{\rm eV})^5$\\
&&~~~~~~~~~~~~~$+4.989108920299513\times10^{-6}({\rm ln}T_{\rm eV})^6$\\
&&~~~~~~~~~~~~~$+5.755614137575758\times10^{-7}({\rm ln}T_{\rm eV})^7$\\
&&~~~~~~~~~~~~~$-1.856767039775261\times10^{-8}({\rm ln}T_{\rm eV})^8$\\
&&~~~~~~~~~~~~~$-3.071135243196595\times10^{-9}({\rm ln}T_{\rm eV})^9]$\\ \\
 
(4)& ${\rm He}^+ + {\rm e}^- \rightarrow {\rm He}+\gamma$ &  $ k_4=1.54\times10^{-9}[1+0.3/{\rm exp}(8.099328789667/T_{\rm eV})]/ $ && cm$^3$ s$^{-1}$\\
&&~~~~~~~$[{\rm exp}(40.49664394833662/T_{\rm eV})\times T_{\rm eV}^{1.5}]$\\
&&~~~~~~~$+3.92\times10^{-13}/T_{\rm eV}^{0.6353} $\\ \\
 
(5)& ${\rm H} + {\rm H}^+ \rightarrow {\rm H_2}^+ +\gamma $ & $k_5=1.85\times10^{-23}T^{1.8}$&$T\le 6.7\times10^3 {\rm K}$  & cm$^3$ s$^{-1}$ \\
&&$   k_5=5.81\times 10^{-16}(T/56200)^{[-0.6657{\rm log}(T/56200)]} $& $T > 6.7\times10^3 {\rm K}  $ \\ \\

(6)& ${\rm H}_2^+ + {\rm H} \rightarrow {\rm H}_2 + {\rm H}^+$ & $k_6=6.0\times10^{-10}$  && cm$^3$ s$^{-1}$\\ \\

(7) &$ {\rm H} + {\rm e}^- \rightarrow {\rm H}^- + \gamma$ & $k_7=6.775\times10^{-15}(T_{\rm eV})^{0.8779} $  && cm$^3$ s$^{-1}$ \\  \\
     
(8) &$ {\rm H} + {\rm H}^- \rightarrow {\rm H}_2 + {\rm e}^-$ & $k_{8}= 1.43\times10^{-9}$ &  $T_{\rm eV} \le 0.1$  & cm$^3$ s$^{-1}$\\
&&$  k_{8}={\rm exp}[-20.06913897587003$ & $T_{\rm eV} > 0.1$\\
&&~~~~~~~~~~~~~$ +0.2289800603272916({\rm ln}T_{\rm eV})$\\
&&~~~~~~~~~~~~~$ +0.03599837721023835({\rm ln}T_{\rm eV})^2 $\\
&&~~~~~~~~~~~~~$-0.004555120027032095({\rm ln}(T_{\rm eV})^3 $\\
&&~~~~~~~~~~~~~$-0.0003105115447124016({\rm ln}(T_{\rm eV})^4$\\
&&~~~~~~~~~~~~~$ +0.0001073294010367247({\rm ln}(T_{\rm eV})^5$\\
&&~~~~~~~~~~~~~$-8.36671960467864\times10^{-6}({\rm ln}T_{\rm eV})^6$\\
&&~~~~~~~~~~~~~$ +2.238306228891639\times10^{-7}({\rm ln}(T_{\rm eV})^7] $\\     \\
 
(9)& ${\rm H}_2^+ + {\rm e}^- \rightarrow 2{\rm H}$ & $k_{9}=1\times10^{-8}$ & $ T\le 617 {\rm K}$ & cm$^3$ s$^{-1}$ \\
&&$k_{9}=1.32\times10^{-6}T^{-0.76}$ &$T>617 {\rm K}$ \\ \\
 
(10) &${\rm H}_2^+ + {\rm H}^- \rightarrow {\rm H}_2 + {\rm H}$ & $k_{10} = 5\times10^{-6} T^{-0.5}$  && cm$^3$ s$^{-1}$\\ \\

(11) &${\rm H}^- + {\rm H}^+ \rightarrow 2{\rm H}$ & $k_{11}=6.5\times10^{-9} T_{\rm eV}^{-0.5}$  && cm$^3$ s$^{-1}$ \\ \\
       
(12)& ${\rm H_2} + {\rm e}^- \rightarrow {\rm H} + {\rm H}^-$ & $k_{12} =2.7\times10^{-8}T^{-1.27}{\rm exp}(-43000/T) $  && cm$^3$ s$^{-1}$  \\ \\
 
(13) & ${\rm H_2} + {\rm H} \rightarrow 3{\rm H}$   &see \citet{1996ApJ...461..265M}  && cm$^3$ s$^{-1}$ \\ \\
 
 (14)   &${\rm H}_2 + {\rm H}_2 \rightarrow {\rm H}_2 + 2{\rm H}$ & $k_{\rm 14,L} = 1.18\times10^{-10}{\rm exp}(-6.95\times10^4/T)$&& cm$^3$ s$^{-1}$ \\
&&$ k_{\rm 14,H}= 8.125\times10^{-8}T^{-1/2}{\rm exp}(-5.2\times10^4/T)[1-{\rm exp}(-6\times10^3/T)]$ \\ 
&&${\rm log}n_{\rm cr,H_2}=4.845-1.3{\rm log}(T/10^4)+1.62[{\rm log}(T/10^4)]^2$ \\
&& $a=(1+n_{\rm H}/n_{\rm cr,H_2})^{-1}$, $k_{14}=k_{\rm 14,H}^{1-a}k_{\rm 14,L}^a$ \\ \\
 
(15) &${\rm H}_2 + {\rm H}^+ \rightarrow {\rm H}_2^+ + {\rm H}$ & $k_{15}={\rm exp}[-24.24914687731536 $  && cm$^3$ s$^{-1}$  \\
&&~~~~~~~~~~~~~~$  +3.400824447095291({\rm ln} T_{\rm eV}) $ & \\
&&~~~~~~~~~~~~~~$ -3.898003964650152({\rm ln}T_{\rm eV})^2 $ & \\
&&~~~~~~~~~~~~~~$ +2.045587822403071({\rm ln}T_{\rm eV})^3 $ & \\
&&~~~~~~~~~~~~~~$-0.5416182856220388({\rm ln}T_{\rm eV})^4 $ & \\
&&~~~~~~~~~~~~~~$+0.0841077503763412({\rm ln}T_{\rm eV})^5$& \\
&&~~~~~~~~~~~~~~$-0.007879026154483455({\rm ln}T_{\rm eV})^6 $ & \\
&&~~~~~~~~~~~~~~$ +0.0004138398421504563({\rm ln}T_{\rm eV})^7 $ & \\
&&~~~~~~~~~~~~~~$-9.36345888928611\times 10^{-6}({\rm ln}T_{\rm eV})^8]  $ & \\ \\ 
 
(16) &  ${\rm H}_2 + {\rm e}^- \rightarrow 2{\rm H}+{\rm e}^-$ & $5.6\times10^{-11}{\rm exp}(-102124/T)T^{0.5}$ && cm$^3$ s$^{-1}$\\ \\   
  
(17) &${\rm H}^- + {\rm e}^- \rightarrow {\rm H}+2{\rm e}^-$ &  $k_{17}={\rm exp}[-18.01849334273 $   && cm$^3$ s$^{-1}$ \\
&&~~~~~~~~~~~~~~$ +2.360852208681({\rm ln}T_{\rm eV}) $ \\
&&~~~~~~~~~~~~~~$-0.2827443061704({\rm ln}T_{\rm eV})^2 $ \\
&&~~~~~~~~~~~~~~$+0.01623316639567({\rm ln}T_{\rm eV})^3 $ \\
&&~~~~~~~~~~~~~~$-0.03365012031362999({\rm ln}T_{\rm eV})^4 $ \\
&&~~~~~~~~~~~~~~$+0.01178329782711({\rm ln}T_{\rm eV})^5$ \\
&&~~~~~~~~~~~~~~$-0.001656194699504({\rm ln}T_{\rm eV})^6$ \\
&&~~~~~~~~~~~~~~$+0.0001068275202678({\rm ln}T_{\rm eV})^7$ \\
&&~~~~~~~~~~~~~~$-2.631285809207\times 10^{-6}({\rm ln}T_{\rm eV})^8]$ \\ \\
 
(18) &${\rm H}^- + {\rm H} \rightarrow 2{\rm H} + {\rm e}^-$ & $k_{18}=2.56\times 10^{-9}T_{\rm eV}^{1.78186}$ & $T_{\rm eV} \le 0.1$  & cm$^3$ s$^{-1}$ \\
&&$k_{18}={\rm exp}[-20.37260896533324 $ & $T_{\rm eV} > 0.1 $ \\
&&~~~~~~~~~~~~~~~$+1.139449335841631({\rm ln}T_{\rm eV}) $ \\
&&~~~~~~~~~~~~~~~$-0.1421013521554148({\rm ln}T_{\rm eV})^2 $ \\ 
&&~~~~~~~~~~~~~~~$+0.00846445538663({\rm ln}T_{\rm eV})^3 $ \\
&&~~~~~~~~~~~~~~~$-0.0014327641212992({\rm ln}T_{\rm eV})^4 $ \\
&&~~~~~~~~~~~~~~~$+0.0002012250284791({\rm ln}T_{\rm eV})^5 $ \\
&&~~~~~~~~~~~~~~~$+0.0000866396324309({\rm ln}T_{\rm eV})^6$ \\
&&~~~~~~~~~~~~~~~$-0.00002585009680264({\rm ln}T_{\rm eV})^7$ \\
&&~~~~~~~~~~~~~~~$+2.4555011970392\times10^{-6}({\rm ln}T_{\rm eV})^8$ \\
&&~~~~~~~~~~~~~~~$-8.06838246118\times10^{-8}({\rm ln}T_{\rm eV})^9]$ \\ \\

(19) &${\rm H}^- + {\rm H}^+ \rightarrow {\rm H}_2^+ + {\rm e}^-$ &    $   k_{19}=4\times10^{-4}T^{-1.4}{\rm exp}(-15100/T) $ & $T \le 10^4 {\rm K}$ & cm$^3$ s$^{-1}$\\
&&$k_{19}=1\times10^{-8}T^{-0.4}$ & $T > 10^4 {\rm K} $ \\ \\
       
(20)& ${\rm H} + {\rm H}+{\rm H} \rightarrow {\rm H}_2 +{\rm H}$ & $k_{20}= 5.5\times10^{-29}T^{-1}   $   && cm$^6$ s$^{-1}$   \\ \\
  
(21) &${\rm H}+{\rm H}+{\rm H}_2 \rightarrow {\rm H}_2 + {\rm H}_2$ & $k_{21}=k_{20}/8$ && cm$^6$ s$^{-1}$\\  \\
  
(22) &$\rm H_2 +\gamma \rightarrow 2{\rm H}$ & see Eq. (\ref{eq_k22_BB}, \ref{eq_k22_GAL} \& \ref{eq_k22_BH})    && s$^{-1}$  \\ \\
  
(23) &${\rm H}^- + \gamma \rightarrow {\rm H} +{\rm e}^-$   &  see Eq. (\ref{eq_k23_BB}, \ref{eq_k23_GAL} \& \ref{eq_k23_BH})  && s$^{-1}$  \\ \\
  
(24) &${\rm H}_2^++\gamma \rightarrow {\rm H} + {\rm H}^+$ & see Eq. (\ref{eq_k24_BB}, \ref{eq_k24_GAL} \& \ref{eq_k24_BH})   && s$^{-1}$  
\enddata

\tablenotetext{a}{$T_{\rm eV}$  and $T$ are temperature in units eV and K respectively. }

\tablenotetext{b}{Reaction (12) is from \citet{2008MNRAS.388.1627G}; reaction (14) is from \citet{2000ApJ...534..809O};  reactions (20) and (21) are from \citet{1983ApJ...271..632P}; 
others are from \citet{2010MNRAS.402.1249S}; original references are found in these cited papers.}  
\label{tb_reactions}
\end{deluxetable}

\LongTables

\begin{deluxetable}{lllll}
\tablecaption{Heating functions used in this work.}
\tablehead{\colhead{Oridinal} &\colhead{Processes} &\colhead{Heating functions\tablenotemark{d}  [erg cm$^{-3}$ s$^{-1}$]}  &  &} 
\startdata

(1)\tablenotemark{e} & 2 body H$_2$ formation& ${\mathcal H}_{2b}=1.6\times10^{-12}\{k_8n_{\rm H^-}n_{\rm H}[3.53(1+n_{\rm cr}/n_{\rm H})^{-1}]+ k_6n_{\rm H_2^+}n_{\rm H}[1.83(1+n_{\rm cr}/n_{\rm H})^{-1}]\} $ \\

(2)& 3 body H$_2$ formation & ${\mathcal H}_{3b}=1.6\times10^{-12}\times4.48(k_{20}n_{\rm H}n_{\rm H}n_{\rm H}+k_{21}n_{\rm H}n_{\rm H}n_{\rm H_2})$ \\

\enddata
\tablenotetext{d}{All from \citet{2006ApJ...652....6Y}, original references are found therein.} 
\tablenotetext{e}{$n_{\rm cr}=10^6T^{-1/2}\left\{1.6y_{\rm H}{\rm exp}\left[-\left(\frac{400}{T}\right)^2\right] +1.4y_{\rm H_2} {\rm exp}\left[-\left(\frac{12000}{T+1200}\right)\right]    \right\}^{-1}$, $y_{\rm H}$ and $y_{\rm H_2}$ are H and H$_2$ number fractions respectively.}
\label{tb_heatings}
\end{deluxetable}

\begin{deluxetable}{lllll}
\tablecaption{Cooling functions used in this work.}
\tablehead{\colhead{Oridinal} &\colhead{Processes} &\colhead{Cooling functions\tablenotemark{f,g}  [erg cm$^{-3}$ s$^{-1}$]}  &  &} 

\startdata

(1) & collisional excitation &$ \Lambda_{\rm H_2,n_{\rm H}\rightarrow0}={\rm dex}[-103.0+97.59{\rm log}T-48.05({\rm log}T)^2+ 10.80({\rm log}T)^3-0.9032({\rm log}T)^4] n_{\rm H} n_{\rm H_2}$ \\ 
&& $\Lambda_{\rm H_2,LTE}=\left\{6.7\times10^{-19}{\rm exp}\left(-\dfrac{5.86}{ T_3}\right)+1.6\times10^{-18}{\rm exp}\left(-\dfrac{11.7}{T_3}\right) \right.$ \\ 
&&~~~~~~~~~~~~~~~~~$\left. +\dfrac{9.5\times10^{-22}T_3^{3.76}}{1+0.12T_3^{2.1}}{\rm exp}\left[-\left(\dfrac{0.13}{T_3}\right)^3\right]+
       3\times10^{-24}{\rm exp}\left(-\dfrac{0.51}{T_3}\right)\right\} n_{\rm H_2}$ \\ 
 &&$\Lambda^{\varA{col-exc}}_{\rm H_2}=\dfrac{\Lambda_{\rm H_2,LTE}} {(1+\Lambda_{\rm H_2,LTE}/\Lambda_{\rm H_2,n_{\rm H} \rightarrow 0})}$ \\ \\

 (2) & collisional dissociation &  $ \Lambda_{\rm H_2}^{\varA{ col-dis}} =7.2\times10^{-12}(k_{13}n_{\rm H}+k_{14}n_{\rm H_2})n_{\rm H_2}$ \\ \\
       
(3)& recombination &$\Lambda_{\rm H}^{\rm rec} =8.7\times 10^{-27}T^{0.5}T_3^{-0.2}(1+T_6^{0.7})^{-1}n_{\rm e}n_{\rm H^+}$ \\ \\
&& $\Lambda_{\rm He}^{\rm rec}=1.55\times10^{-26}T^{0.3647} n_{\rm e} n_{\rm He^+}$   \\ \\

(4)& collisional ionization & $\Lambda_{\rm H}^\varA{col-ion}=  1.27\times10^{-21}T^{1/2}(1+T_5^{0.5} )^{-1} {\rm exp}\left(-\dfrac{157809.1}{T}\right)n_{\rm e}n_{\rm H} $  \\ 
&&$\Lambda_{\rm He}^\varA{col-ion}=   9.38\times 10^{-22}T^{0.5}(1+T_5^{0.5})^{-1}{\rm  exp}\left(-\dfrac{285335.4}{T}\right) n_{\rm e}n_{\rm He}  $    \\ \\                           

(5) &collisional excitation  & $\Lambda_{\rm H}^\varA{col-exc} = 7.5\times10^{-19}(1+T_5^{0.5} )^{-1}{\rm exp}\left(-\dfrac{118348}{T}\right) n_{\rm e}n_{\rm H}  $    \\ 
&& $\Lambda_{\rm He}^\varA{col-exc} = 9.1\times10^{-27}T^{-0.1687}(1+T_5^{0.5} )^{-1}{\rm exp}\left(-\dfrac{13179}{T}\right)n_{\rm e}^2n_{\rm He^+}$ \\ 
&&$\Lambda_{\rm He^+}^\varA{col-exc} =  5.54\times10^{-17}T^{-0.397}(1+T_5^{0.5} )^{-1}{\rm exp}\left(-\dfrac{473638}{T}\right)n_{\rm e}n_{\rm He^+} $ \\ \\

(6) &Bremsstrahlung & $\Lambda^{\rm Bre} = 1.42\times 10^{-27}T^{0.5}(n_{\rm H^+}+n_{\rm He^+})n_{\rm e}$   \\ \\

(7)& Compton & $\Lambda^{\rm com} = 1.017\times10^{-37}T_{\rm CMB}^4(T-T_{\rm CMB})n_{\rm e} $        

\enddata
\tablenotetext{f}{$T_3 = T/10^3 ~{\rm K}$, $T_5 = T/10^5~{\rm K}$, $T_6 = T/10^6~{\rm K}$} 
\tablenotetext{g}{Function (1) is from \citet{1998A&A...335..403G} \& \citet{1979ApJS...41..555H};  function (2) from  \citet{2007ApJ...666....1G}; 
functions (3) - (5) are from \citet{2003MNRAS.345..379M}; functions (6) - (7) are from \citet{2003MNRAS.345..379M}; 
original references are found in these cited papers.}
\label{tb_coolings}
\end{deluxetable}

\end{document}